\documentclass[pra,showpacs,amsmath,amssymb,twocolumn]{revtex4-1}
\usepackage{bm}
 \usepackage{graphicx}
 \usepackage[bookmarks=false]{hyperref}

\newcommand{\bra}[1]{\langle #1|}
\newcommand{\ket}[1]{|#1\rangle}
\newcommand{\braket}[1]{\langle #1 \rangle}

\def\dd{\mathrm{d}}
\def\ee{\mathrm{e}}
\def\ii{\mathrm{i}}
\def\vnabla{\bm{\nabla}}
\def\grad{\vnabla}
\def\div{\vnabla\cdot}
\def\rot{\vnabla\times}

\def\ddt#1{\frac{\dd #1}{\dd t}}
\def\ddtt#1{\frac{\dd^2 #1}{\dd t^2}}
\def\ddx#1{\frac{\partial #1}{\partial x}}
\def\ddxx#1{\frac{\partial^2 #1}{\partial x^2}}
\def\ddz#1{\frac{\partial #1}{\partial z}}

\def\const{\text{const.}}
\def\Hc{\text{H.c.}}
\def\pbar{\partial\llap{--}}

\def\dz{\varDelta z}

\def\modu{\zeta}

\def\CT{C_{\text{T}}}
\def\LT{L_{\text{T}}}
\def\ZT{Z_{\text{T}}}
\def\CC{C_{\text{c}}}
\def\CCR{C_{\text{c}}'}
\def\CCCR{C_{\text{c}}''}
\def\wz{\omega_0}

\def\wx{\omega_\text{x}}
\def\CR{C_\text{R}}
\def\LR{L_\text{R}}
\def\ZR{Z_\text{R}}
\def\dx{\varDelta x}
\def\cavlen{d}
\def\len{\ell}

\def\CJ{C_{\text{J}}}
\def\EJ{E_{\text{J}}}
\def\ECP{E_{\text{CP}}}

\def\rabi{g}
\def\grabi{\tilde{g}}
\def\kappabare{\kappa_{\text{bare}}}
\def\kappaLC{\kappa}
\def\kappaLCz{\kappa_{\text{LC}}}
\def\kappaTLR{\kappa^{\text{TLR}}}

\def\kappaFPz{\kappa^{\text{FP}}_0}

\def\LL{\mathcal{L}}
\def\Lz{\LL_0}
\def\LA{\LL_M}

\def\HH{\mathcal{H}}
\def\Hz{\HH_0}
\def\HE{\HH_{\text{env}}}
\def\HSE{\HH_{\text{SEC}}}
\def\HLC{\HH_{\text{LC}}}

\def\oHH{\hat{\mathcal{H}}}

\def\oHz{\oHH_0}
\def\oHE{\oHH_{\text{env}}}
\def\oHSE{\oHH_{\text{SEC}}}
\def\oHmat{\oHH_{\text{mat}}}
\def\oHLC{\oHH_{\text{LC}}}

\def\oa{\hat{a}}
\def\oain{\hat{a}_{\text{in}}}

\def\oaout{\hat{a}_{\text{out}}}
\def\oad{\hat{a}^{\dagger}}
\def\ob{\hat{b}}
\def\obd{\hat{b}^{\dagger}}

\def\op{\hat{p}}
\def\opd{\hat{p}^{\dagger}}
\def\oQ{\hat{Q}}
\def\oPhi{\hat{\varPhi}}
\def\oPhiin{\hat{\varPhi}_{\text{in}}}
\def\oPhiout{\hat{\varPhi}_{\text{out}}}
\def\oq{\hat{q}}
\def\ophi{\hat{\phi}}
\def\oalpha{\hat{\alpha}}
\def\oalphad{\hat{\alpha}^{\dagger}}
\def\osigma{\hat{\sigma}}
\def\osigmad{\hat{\sigma}^{\dagger}}
\def\opsi{\hat{\psi}}
\def\orho{\hat{\rho}}
\def\oU{\hat{U}}
\def\oUd{\hat{U}^{\dagger}}
\def\oX{\hat{X}}
\def\ox{\hat{x}}
\def\oxd{\hat{x}^{\dagger}}

\def\oO{\hat{O}}
\def\oS{\hat{S}}
\def\os{\hat{s}}
\def\osd{\hat{s}^{\dagger}}

\def\LDH{\LL_{DH}}
\def\LDHM{\LL_{M-DH}}
\def\LEB{\LL_{EB}}
\def\LEBM{\LL_{M-EB}}

\def\diez{\varepsilon_0}
\def\diebg{\varepsilon_{\text{bg}}}
\def\muz{\mu_0}
\def\scp{\varphi}

\def\vr{\bm{r}}
\def\vA{\bm{A}}
\def\vE{\bm{E}}
\def\vET{\bm{E}_{\perp}}
\def\vEL{\bm{E}_{\parallel}}
\def\vB{\bm{B}}
\def\vD{\bm{D}}
\def\vH{\bm{H}}
\def\vHT{\bm{H}_{\perp}}
\def\vHL{\bm{H}_{\parallel}}
\def\vP{\bm{P}}

\def\vM{\bm{M}}

\def\vY{\bm{Y}}

\def\YY{Y}

\def\oY{\hat{Y}}

\def\oE{\hat{E}}
\def\oB{\hat{B}}

\def\oH{\hat{H}}
\def\oHin{\hat{H}_{\text{in}}}
\def\oHout{\hat{H}_{\text{out}}}

\begin{document}

%\preprint{APS/123-QED}

\title{A recipe for Hamiltonian of system-environment coupling
applicable to ultrastrong light-matter interaction regime}

\author{Motoaki Bamba}
\altaffiliation{E-mail: bamba@acty.phys.sci.osaka-u.ac.jp}
\affiliation{Department of Physics, Osaka University, 1-1 Machikaneyama, Toyonaka, Osaka 560-0043, Japan}
\author{Tetsuo Ogawa}
\affiliation{Department of Physics, Osaka University, 1-1 Machikaneyama, Toyonaka, Osaka 560-0043, Japan}
\date{\today}

\date{\today}

\begin{abstract}
When the light interacts with matters in a lossy cavity,
in the standard cavity quantum electrodynamics,
the dissipation of cavity fields is characterized simply by the strengths of the two couplings:
the light-matter interaction and the system-environment coupling through the cavity mirror.
However, in the ultrastrong light-matter interaction regime,
the dissipation depends also on whether the two couplings
are mediated by the electric field or the magnetic one
(capacitive or inductive in superconducting circuits).
Even if we know correctly
the microscopic mechanism (Lagrangian) of the system-environment coupling,
the coupling Hamiltonian itself is in principle modified
due to the ultrastrong interaction in the cavity.
In this paper, we show a recipe for deriving a general expression
of the Hamiltonian of the system-environment coupling,
which is applicable even in the ultrastrong light-matter interaction regime
in the good-cavity and independent-transition limit.
\end{abstract}


% 42.50.Ct	Quantum description of interaction of light and matter; related experiments
% 85.25.Am	Superconducting device characterization, design, and modeling
% 42.50.Pq	Cavity quantum electrodynamics; micromasers
% 03.65.Yz	Decoherence; open systems; quantum statistical methods

\maketitle
\section{Introduction}
When the electromagnetic fields are confined in a cavity consisting of mirrors
with high reflectivities, we can energetically identify the cavity modes
while they have inevitable broadenings due to the loss through the mirrors.
The loss is usually characterized by a loss rate $\kappabare$
that is determined by the reflectivity of the mirror
(or the coupling strength between the cavity mode and its environment)
and the density of states of the environment.
In the standard cavity quantum electrodynamics (QED) \cite{gardiner04,walls08},
the Hamiltonian of the system-environment coupling (SEC) is simply supposed as
\begin{equation} \label{eq:oHSE_standard} % !!!!!!!!!!!!!!!!!!!!!!!!!!!!!!!!!!
\oHSE^{\text{standard}} = 
\int_0^{\infty}\dd\omega\
  \ii\hbar\sqrt{\frac{\kappabare(\omega)}{2\pi}}
  \left[ \oalphad(\omega)\oa - \oad\oalpha(\omega) \right].
\end{equation}
Here, $\oa$ is the annihilation operator of a photon in the cavity mode
with eigen-frequency $\wz$,
and $\oalpha(\omega)$ is the one in the environment with frequency $\omega$.
They satisfy $[\oa, \oad] = 1$, 
$[\oalpha(\omega), \oalphad(\omega')] = \delta(\omega-\omega')$,
and $[\oa, \oa] = [\oalpha(\omega), \oalpha(\omega')]
= [\oa, \oalpha(\omega)] = [\oa, \oalphad(\omega)] = 0$.
In this standard expression,
photons pass through the cavity mirror
with conserving the number of photons,
and we cannot catch the information of
whether the SEC is through the electric field or the magnetic one
by watching only the above expression.
This question has tiny meaning
if Eq.~\eqref{eq:oHSE_standard} is well justified \cite{Nakatani2010JPSJ},
and we can discuss simply the dissipative motion of ``photons''
without considering the electric and magnetic fields explicitly.

When the cavity embeds matters with excitations interacting with the electromagnetic fields,
the Hamiltonian of the cavity system is generally expressed as
\footnote{$(\oa+\oad)^2$ or $S_x{}^2$ term is renormalized to $\wz$ or $\wx$}
\begin{equation} \label{eq:oHz_noRWA} % !!!!!!!!!!!!!!!!!!!!!!!!!!!!!!!!!!!!!!
\oHz = \hbar\wz\oad\oa + \hbar\rabi(\oa+\oad)\oS_x + \oHmat.
\end{equation}
Here, $\oHmat$ is the Hamiltonian of the matters,
$\oS_x$ is the non-dimensional operator that annihilates or creates an excitation in matters,
and $\rabi$ is the strength of the light-matter interaction
(vacuum Rabi splitting).
If $\rabi$ is small enough compared to $\wz$
and to the transition frequency $\wx$ of matters ($\wz, \wx \gg \rabi$),
the rotating-wave approximation (RWA) can be applied to the light-matter interaction,
and then $\oHz$ can be rewritten with the annihilation operator $\os$
and creation one $\osd$ of an excitation in matters ($\oS_x = \os + \osd$) as
\begin{equation} \label{eq:oHz_RWA} % !!!!!!!!!!!!!!!!!!!!!!!!!!!!!!!!!!!!!!!!
\oHz \simeq 
\oHz^{\text{RWA}} = \hbar\wz\oad\oa + \hbar\rabi(\osd\oa+\oad\os) + \oHmat.
\end{equation}
Under the RWA, the total number of photons and excitations is conserved
in the process of the light-matter interaction,
and at the same time 
the standard SEC Hamiltonian \eqref{eq:oHSE_standard} can be justified
\cite{gardiner04,Breuer2006,Beaudoin2011PRA,Bamba2013MBC}.
In most of the studies of cavity QED,
Eqs.~\eqref{eq:oHz_RWA} and \eqref{eq:oHSE_standard}
are supposed as the standard Hamiltonians of the cavity system with light-matter interaction
and of the SEC, respectively.
When the loss rate of cavity fields or that of excitations
is higher than the light-matter interaction strength
($\kappabare > \rabi$), it is called the weak interaction (coupling) regime,
and the opposite case $\wz, \wx \gg \rabi > \kappabare$
is called the (normally) strong interaction (coupling) one.

However, when the electromagnetic fields ultrastrongly interact with matters
($\rabi \gtrsim \wz,\wx$) \cite{Ciuti2005PRB,Devoret2007AP},
the RWA cannot be applied to the light-matter interaction,
and we must explicitly consider the counter-rotating terms
(in the basis of photons and excitations) $\oa\os$ and $\oad\osd$ in Eq.~\eqref{eq:oHz_noRWA}.
The ultrastrong interaction has been realized with a great effort recently
by the electric dipole transitions
in subbands of semiconductor quantum wells \cite{Gunter2009N,Anappara2009PRB,Todorov2009PRL,Todorov2010PRL,Todorov2012PRB,Porer2012PRB},
molecular materials \cite{Schwartz2011PRL},
and two-dimensional electron gas \cite{Scalari2012S}.
Further, in superconducting circuits,
which has a good correspondence with the cavity QED
\cite{Blais2004PRA,Devoret2007AP},
flux qubits also exhibit the ultrastrong interaction
with fields in transmission line resonator \cite{Niemczyk2010NP,Fedorov2010PRL,Forn-Diaz2010PRL}.

In the ultrastrong light-matter interaction regime,
the standard expression \eqref{eq:oHSE_standard} of SEC is failed
\cite{Beaudoin2011PRA,Bamba2012DissipationUSC,Bamba2013MBC},
and the counter-rotating terms $\oa\oalpha(\omega)$ and $\oalphad(\omega)\oad$
must also be considered for the SEC
(the reason will be explained later).
Instead of Eq.~\eqref{eq:oHSE_standard},
we can suppose intuitively the following two SEC Hamiltonians:
\begin{subequations} \label{eq:oHSE_pm} % !!!!!!!!!!!!!!!!!!!!!!!!!!!!!!!!!!!!
\begin{align}
\oHSE^+
& = \int_0^{\infty}\dd\omega\
    \ii\hbar\sqrt{\frac{\kappabare(\omega)}{2\pi}}(\oa+\oad)
    \left[ \oalphad(\omega) - \oalpha(\omega) \right], \label{eq:oHSE_+} \\ % !!
\oHSE^-
& = \int_0^{\infty}\dd\omega\
    \ii\hbar\sqrt{\frac{\kappabare(\omega)}{2\pi}}[\ii(\oa-\oad)]
    \left[ \oalphad(\omega) - \oalpha(\omega) \right]. \label{eq:oHSE_-} % !!!
\end{align}
\end{subequations}
These expressions will be obtained in this paper
by the straightforward calculation from Lagrangians.
Since the light-matter interaction is proportional to $\oa+\oad$
[not to $\ii(\oa-\oad)$] in Eq.~\eqref{eq:oHz_noRWA},
the dissipative motion of the cavity fields in principle depends
on whether the SEC is expressed as $\oHSE^+$ or $\oHSE^-$.
Then, in the ultrastrong light-matter interaction regime,
we face the ambiguity for choosing $\oHSE^{\pm}$
(or the sign in the light-matter interaction),
by which the dissipation of the cavity fields is also characterized
not only by the strengths of the two couplings ($\rabi$ and $\kappabare$)
and the detuning between $\wz$ and $\wx$.
This fact will be demonstrated also in this paper.
In other words,
the ``photon'' picture is no longer suitable,
and we must correctly understand whether the two couplings
(light-matter interaction and SEC)
are mediated by the electric field or the magnetic one.
They are basically determined by the detailed mechanisms
of the light-matter interaction and of the SEC in the system to be supposed
(described by Lagrangians),
and we can no longer discuss the dissipation and detection
in the ultrastrong light-matter interaction regime
without avoiding this ambiguity.

The theoretical treatment of the SEC in the ultrastrong light-matter interaction regime
has been discussed
in the formalism of quantum Langevin equation \cite{Ciuti2006PRA,Liberato2007PRL}
and of master equation \cite{DeLiberato2009PRA,Beaudoin2011PRA}.
The photon measurement has also been discussed \cite{Ridolfo2012PRL},
and many theoretical proposals were raised
\cite{Liberato2007PRL,Ridolfo2012PRL,Ridolfo2013PRL,Stassi2013PRL,Liberato2013a}.
However, in these theoretical works, the SEC Hamiltonians were simply given
without the discussion whether it is mediated by the electric field
or the magnetic one (capacitive or inductive in circuits).
The derivation of the SEC Hamiltonians has been discussed
by considering explicitly the boundary conditions of the cavity fields
\cite{Knoll1991PRA,gruner96mar,Dutra2000JOB,Dutra2000PRA,Dalton2001PRA,Hackenbroich2002PRL,Viviescas2003PRA,Khanbekyan2005PRA,Johansson2010PRA},
whereas empty cavities were mainly supposed.
In our previous work \cite{Bamba2013MBC},
the SEC Hamiltonian is derived based on the Maxwell's boundary conditions
at the mirror of the cavity that is filled by a medium with excitations
ultrastrongly interacting with the electromagnetic fields.
However, the derivation is applicable to only the systems with simple bosonic excitations in matters,
and a recipe for deriving the SEC Hamiltonian that is independent of the detail of matters is still desired.
In this paper, we try to derive the SEC Hamiltonian
starting from the Lagrangian describing the detailed mechanisms of the SEC
while remaining the matter system as a black box as far as possible.
The SEC is explicitly discussed for two systems:
superconducting circuits and a Fabry-Perot cavity.

We first review the treatment of the SEC in Sec.~\ref{sec:review}.
The Langevin and master equations are derived for given SEC Hamiltonians
\eqref{eq:oHSE_pm}, and the input-output relation is also derived.
The straightforward derivation of the SEC Hamiltonians is performed
in Sec.~\ref{sec:straight}.
We will find that,
in the straightforward calculation from the Lagrangians,
the SEC Hamiltonians are in principle modified
even in presence of the ultrastrong light-matter interaction.
In other words, the SEC Hamiltonians cannot be determined
by the complete knowledge of the cavity systems in principle.
In some cases, the SEC Hamiltonians cannot be obtained
in such simple forms as Eqs.~\eqref{eq:oHSE_pm}.
However, in the good-cavity and independent-transition limit
(definitions will be explained in Sec.~\ref{sec:review}),
the SEC Hamiltonians can be derived as Eqs.~\eqref{eq:oHSE_pm},
and the expressions are not modified by the presence of the ultrastrong
light-matter interaction.
The recipe of the derivation will be shown in Sec.~\ref{sec:derive_SEC}.
The discussions in Secs.~\ref{sec:straight} and \ref{sec:derive_SEC}
will be performed by considering explicitly superconducting circuits
consisting of a LC-resonator as depicted in Fig.~\ref{fig:1}
(the case of a transmission line resonator is discussed
in App.~\ref{sec:TLR-TL}).
The derivation of the SEC Hamiltonian
of a Fabry-Perot cavity for the electromagnetic fields
will be shown in Sec.~\ref{sec:FP}.
The discussion in this paper will be summarized in Sec.~\ref{sec:summary}.

\section{Reviewing treatment of SEC} \label{sec:review}
When the total Hamiltonian $\oHH = \oHz + \oHSE^{\pm} + \oHE$ is already given,
we can discuss the dissipation of the cavity system $\oHz$
based on the well-established frameworks \cite{gardiner04,Breuer2006}
even in the ultrastrong interaction regime.
Here, the Hamiltonian of the environment is simply supposed as
\begin{equation}
\oHE = \int_0^{\infty}\dd\omega\ \hbar\omega\oalphad(\omega)\oalpha(\omega),
\end{equation}
and the annihilation and creation operators satisfy
\begin{subequations} \label{eq:[alpha,alpha]} % !!!!!!!!!!!!!!!!!!!!!!!!!!!!!!
\begin{align}
\left[\oalpha(\omega), \oalphad(\omega')\right] & = \delta(\omega-\omega'), \\
\left[\oalpha(\omega), \oalpha(\omega')\right] & = 0.
\end{align}
\end{subequations}
We first diagonalize the cavity system including the light-matter interaction as
\begin{equation}
\oHz = \sum_{\mu} \hbar\omega_{\mu}\ket{\mu}\bra{\mu},
\end{equation}
where $\ket{\mu}$ is an eigen-state and $\omega_{\mu}$ is its eigen-frequency.
As mentioned in Ref.~\cite{Beaudoin2011PRA}
for the ultrastrong light-matter interaction regime,
in order to guarantee the decay of the cavity system
to its ground state in the environment at zero temperature
\cite{Bamba2012DissipationUSC,Bamba2013MBC},
we should neglect the fast-rotating (counter-rotating) terms
in the SEC Hamiltonian \eqref{eq:oHSE_pm} 
based on the eigen-states of the cavity system as
\begin{equation} \label{eq:oHSE_pre} % !!!!!!!!!!!!!!!!!!!!!!!!!!!!!!!!!!!!!!!
\oHSE^{\pm}
\simeq \int_0^{\infty}\dd\omega\
    \ii\hbar\sqrt{\frac{\kappabare(\omega)}{2\pi}}
    \left[ \oalphad(\omega) \ox_{\pm} - \oxd_{\pm}\oalpha(\omega) \right].
\end{equation}
Here, we denote the cavity fields as
\begin{subequations}
\begin{align}
\oX_+ & = \oa + \oad, \\
\oX_- & = \ii(\oa - \oad),
\end{align}
\end{subequations}
and $\ox_{\pm}$ represent the lowering part of them:
\begin{equation}
\ox_{\pm} = \sum_{\mu,\nu>\mu} \ket{\mu}\braket{\mu|\oX_{\pm}|\nu}\bra{\nu}.
\end{equation}
In the weak and normally strong light-matter interaction regimes,
the lowering and raising operators reduce to $\ox_{\pm} = \oa$ and $\oxd_{\pm} = \oad$,
respectively.
Then, the SEC Hamiltonian \eqref{eq:oHSE_pre}
is rewritten to the standard one \eqref{eq:oHSE_standard}.
However, in the ultrastrong light-matter interaction regime,
we generally get $\ox_{\pm} \neq \oa$,
because the total number of photons and excitations is no longer conserved
\cite{Ciuti2005PRB}.
Here, we suppose that $\oX_{\pm}$ does not have diagonal elements for simplicity,
and the original fields are represented as $\oX_{\pm} = \ox_{\pm} + \oxd_{\pm}$,
whereas the diagonal elements cause the pure dephasing in general \cite{Beaudoin2011PRA}.
In the derivation of the master equation,
the importance of the usage of true eigen-states of the cavity system
(called the spectral decomposition in Ref.~\cite{Breuer2006})
has been discussed even for the normally strong light-matter interaction regime
\cite{Carmichael1973JPA,Carmichael1974PRA,Scala2007PRA,Scala2007JPA,Fleming2010JPA},
and the approximation from Eq.~\eqref{eq:oHSE_pm} to Eq.~\eqref{eq:oHSE_pre}
is called the pre-trace RWA in Ref.~\cite{Fleming2010JPA}.
This is because the counter-rotating terms $\oalpha(\omega)\ox_{\pm}$
and $\oalphad(\omega)\oxd_{\pm}$ (in the basis of true eigen-states)
oscillate rapidly compared to the remaining two terms
appearing in Eq.~\eqref{eq:oHSE_pre}.
This approximation is justified 
when the environment is large enough compared to the cavity system
and in the good-cavity limit,
i.e., the loss rate is small enough compared to the transition frequency of the cavity system:
$\kappabare \ll \wz, \wx$
[more precisely $\kappa_{\nu,\mu} < \omega_{\nu,\mu}$
for loss rate $\kappa_{\nu,\mu}$ \eqref{eq:kappa_nu_mu}
relevant to the transition of interest with frequency
$\omega_{\nu,\mu} =  \omega_{\nu} - \omega_{\mu}$ ($\nu > \mu$)].
Under these assumptions,
we can suppose that the state of the environment is modified only slightly
by the coupling with the cavity system,
and the energy loss to the environment no longer return to the cavity.
Then, only the two co-rotating terms that conserve the energy
can survive in Eq.~\eqref{eq:oHSE_pre}.
If we want to discuss the Lamb shift due to the SEC correctly,
we should remain the counter-rotating terms,
and Eqs.~\eqref{eq:oHSE_pm} should be used without applying the pre-trace RWA.

We simply suppose that the fields in the environment have no coherence
and distributed as
\begin{subequations} \label{eq:distribution_alpha} % !!!!!!!!!!!!!!!!!!!!!!!!!
\begin{align}
\braket{\oalphad(\omega)\oalpha(\omega')} & = \delta(\omega-\omega') n(\omega), \\
\braket{\oalpha(\omega)\oalphad(\omega')} & = \delta(\omega-\omega') [n(\omega) + 1], \\
\braket{\oalpha(\omega)\oalpha(\omega')} & = 0.
\end{align}
\end{subequations}
From the approximated SEC Hamiltonian \eqref{eq:oHSE_pre},
the master equation for the reduced density operator $\orho$
representing the cavity system
is derived under the Born-Markov approximation in the Schr\"odinger picture as
\cite{gardiner04,Breuer2006}
\begin{widetext}
\begin{align} \label{eq:master_BornMarkov} % !!!!!!!!!!!!!!!!!!!!!!!!!!!!!!!!!
\ddt{}\orho(t)
& = \frac{1}{\ii\hbar}[ \orho(t), \oHz ]
+ \sum_{\mu,\nu>\mu} \frac{\kappabare(\omega_{\nu,\mu})}{2}\left\{
    n(\omega_{\nu,\mu}) \left[
      \oxd_{\pm}\orho(t)\ox_{\pm}^{\mu,\nu}
    + \{\ox_{\pm}^{\mu,\nu}\}^{\dagger}\orho(t)\ox_{\pm}
    - \ox_{\pm}\{\ox_{\pm}^{\mu,\nu}\}^{\dagger}\orho(t)
    - \orho(t)\ox_{\pm}^{\mu,\nu}\oxd_{\pm}
    \right]
  \right.
\nonumber \\ & \quad
  \left.
  + [n(\omega_{\nu,\mu})+1] \left[
      \ox_{\pm}\orho(t)\{\ox_{\pm}^{\mu,\nu}\}^{\dagger}
    + \ox_{\pm}^{\mu,\nu}\orho(t)\oxd_{\pm}
    - \oxd_{\pm}\ox_{\pm}^{\mu,\nu}\orho(t)
    - \orho(t)\{\ox_{\pm}^{\mu,\nu}\}^{\dagger}\ox_{\pm}
    \right]
\right\},
\end{align}
where
$\ox_{\pm}^{\mu,\nu} = \ket{\mu}\braket{\mu|\oX_{\pm}|\nu}\bra{\nu}$.
Further, we suppose that the loss rate is low enough compared to
the difference of transitions of interest as
$\kappabare \ll |\omega_{\mu,\nu} - \omega_{\mu',\nu'}|$
(more precisely $\kappa_{\nu,\mu}, \kappa_{\nu',\mu'} \ll |\omega_{\mu,\nu} - \omega_{\mu',\nu'}|$)
and all the transitions of interest can be well identified
(we call this situation as ``independent-transition limit'' in this paper).
In this situation, we can also neglect the fast-oscillating terms
involving the different transitions $\{\mu,\nu\} \neq \{\mu',\nu'\}$,
which is called the post-trace RWA in Ref.~\cite{Fleming2010JPA},
and then the master equation is finally approximated as
\cite{Breuer2006,Beaudoin2011PRA}
\begin{align}
\ddt{}\orho(t)
& = \frac{1}{\ii\hbar}[ \orho(t), \oHz ]
+ \sum_{\mu,\nu>\mu} \frac{\kappabare(\omega_{\nu,\mu})|\braket{\mu|\oX_{\pm}|\nu}|^2}{2}\left\{
    n(\omega_{\nu,\mu}) \left[
      2\osigmad_{\mu,\nu}\orho(t)\osigma_{\mu,\nu}
    - \osigma_{\mu,\nu}\osigmad_{\mu,\nu}\orho(t)
    - \orho(t)\osigma_{\mu,\nu}\osigmad_{\mu,\nu}
    \right]
  \right.
\nonumber \\ & \quad
  \left.
  + [n(\omega_{\nu,\mu})+1] \left[
      2\osigma_{\mu,\nu}\orho(t)\osigmad_{\mu,\nu}
    - \osigmad_{\mu,\nu}\osigma_{\mu,\nu}\orho(t)
    - \orho(t)\osigmad_{\mu,\nu}\osigma_{\mu,\nu}
    \right]
\right\},
\end{align}
where $\osigma_{\mu,\nu} = \ket{\mu}\bra{\nu}$.
In this way, the loss rate of each transition is modulated
by the matrix element $|\braket{\mu|\oX_{\pm}|\nu}|^2$
of the cavity field $\oX_{\pm}$ involving the SEC:
\begin{equation} \label{eq:kappa_nu_mu}
\kappa_{\nu,\mu} = \kappabare(\omega_{\nu,\mu})|\braket{\mu|\oX_{\pm}|\nu}|^2.
\end{equation}
In the same manner, we can also treat the dissipation of matters
by considering the coupling between fields of excitations in matters
and its environment including the pure dephasing \cite{Beaudoin2011PRA}.

On the other hand, from the approximated SEC Hamiltonian \eqref{eq:oHSE_pre},
we can also derive the quantum Langevin equation for system operator $\oO$
under the Born-Markov approximation in the Heisenberg picture as \cite{gardiner04}
\begin{align} \label{eq:Langevin_BornMarkov} % !!!!!!!!!!!!!!!!!!!!!!!!!!!!!!!!!
\ddt{}\oO & = \frac{1}{\ii\hbar}\left[ \oO, \oHz \right]
- \left[ \oO, \oxd_{\pm} \right]\int_0^{\infty}\dd\omega\ \ee^{-\ii\omega t}
  \left[ \frac{\kappabare(\omega)}{2}\ox_{\pm}(\omega) + \sqrt{\kappabare(\omega)}\oain(\omega) \right]
\nonumber \\ & \quad
+ \int_0^{\infty}\dd\omega\ \ee^{\ii\omega t}
  \left[ \frac{\kappabare(\omega)}{2}\ox_{\pm}(\omega) + \sqrt{\kappabare(\omega)}\oain(\omega) \right]^{\dagger}
  \left[ \oO, \ox_{\pm} \right],
\end{align}
\end{widetext}
where the Fourier transform is defined as
\begin{subequations}
\begin{align}
\ox(t)
& = \int_{-\infty}^{\infty}\dd\omega\ \ee^{-\ii\omega t} \ox(\omega), \\
\ox(\omega)
& = \frac{1}{2\pi}\int_{-\infty}^{\infty}\dd t\ \ee^{\ii\omega t} \ox(t).
\end{align}
\end{subequations}
The input operator $\oain(t)$ is defined as
\begin{subequations}
\begin{align}
\oain(t) & = \frac{1}{\sqrt{2\pi}}\int_0^{\infty}\dd\omega\ \ee^{-\ii\omega(t-t_0)} \oalpha(\omega,t_0), \\
\oain(\omega) & = \frac{1}{\sqrt{2\pi}} \ee^{\ii\omega t_0} \oalpha(\omega,t_0) \quad \text{for $\omega > 0$},
\end{align}
\end{subequations}
where $t_0$ is the switch-on time of the SEC \cite{gardiner04},
and $\oalpha(\omega,t_0)$ satisfies the equal-time commutation relation
\eqref{eq:[alpha,alpha]} and also show the distribution of the environment
\eqref{eq:distribution_alpha}.
Obeying the procedure in Ref.~\cite{gardiner04},
the input-output relation is also derived as \cite{Ridolfo2012PRL}
\begin{equation} \label{eq:input-output_X} % !!!!!!!!!!!!!!!!!!!!!!!!!!!!!!!!!
\oaout(\omega) = \oain(\omega) + \sqrt{\kappabare(\omega)}\ox_{\pm}(\omega).
\end{equation}
For calculating quantities corresponding to measurements by photon detectors,
we should perform the time- and normal-ordering of the operators \cite{Ridolfo2012PRL}
in the basis of true eigen-states of cavity system,
as usually performed in quantum optics \cite{gardiner04}.
In Ref.~\cite{Ciuti2006PRA}, the standard SEC Hamiltonian \eqref{eq:oHSE_standard} was used
for deriving the quantum Langevin equation and the input-output relation.
Even in this treatment, we can obtain the vacuum output from the vacuum input,
but the cavity itself is excited even by the environment at zero temperature [$n(\omega) = 0$]
\cite{Bamba2012DissipationUSC}.
We can avoid this problem by using the SEC Hamiltonian \eqref{eq:oHSE_pre}
with the pre-trace RWA (in the basis of true eigen-states).
In the independent-transition limit,
the quantum Langevin equation is rewritten
under the post-trace RWA (neglecting the coupling with other transitions) as
\begin{subequations}
\begin{align}
\ddt{}\osigma_{\mu,\nu}
& = \left( - \ii\omega_{\nu,\mu} - \frac{\kappabare(\omega_{\nu,\mu})|\braket{\mu|\oX_{\pm}|\nu}|^2}{2} \right)
    \osigma_{\mu,\nu}
\nonumber \\ & \quad
  - \int_0^{\infty}\dd\omega\ \ee^{-\ii\omega t} \sqrt{\kappabare(\omega)}\braket{\nu|\oX_{\pm}|\mu}\oain(\omega), \\
\osigma_{\mu,\nu}(\omega)
& = \frac{-\ii\sqrt{\kappabare(\omega)}\braket{\nu|\oX_{\pm}|\mu}}{\omega-\omega_{\nu,\mu}+\ii\kappabare(\omega)|\braket{\mu|\oX_{\pm}|\nu}|^2}.
\end{align}
\end{subequations}
Substituting this into the input-output relation \eqref{eq:input-output_X},
we get
\begin{equation}
\oaout(\omega)
= \frac{\omega-\omega_{\nu,\mu}-\ii\kappabare(\omega)|\braket{\mu|\oX_{\pm}|\nu}|^2/2}
       {\omega-\omega_{\nu,\mu}+\ii\kappabare(\omega)|\braket{\mu|\oX_{\pm}|\nu}|^2/2}
  \oain(\omega).
\end{equation}
In this way, we can also check the modulation of the loss rates of the transitions
by the matrix element $|\braket{\mu|\oX_{\pm}|\nu}|^2$.

The above master equation \eqref{eq:master_BornMarkov}
and the quantum Langevin one \eqref{eq:Langevin_BornMarkov} are derived
from the given SEC Hamiltonian $\oHSE^{\pm}$ \eqref{eq:oHSE_pm}.
However, there still remains the ambiguity for choosing $\oX_+ = \oa + \oad$
or $\oX_- = \ii(\oa-\oad)$, which is in principle determined by the
mechanism of the confinement and loss of the cavity fields.
In other words, we get an additional degree of freedom for controlling
the dissipation of the cavity system in the ultrastrong light-matter interaction regime.
In our previous work \cite{Bamba2013MBC},
we supposed a cavity structure consisting of a perfect mirror and a non-perfect
thin mirror as depicted in Fig.~\ref{fig:3}.
Supposing that the inside of the cavity is filled by a medium
with simple bosonic excitations, we successfully derived the SEC Hamiltonian
from the Maxwell's boundary conditions at the non-perfect mirror.
It is expressed as
\begin{equation} \label{eq:oHSE_MBC} % !!!!!!!!!!!!!!!!!!!!!!!!!!!!!!!!!!!!!!!
\oHSE^{\text{MBC}}
= \sum_j \int_0^{\infty}\dd\omega\
  \ii\hbar\sqrt{\frac{\kappa_{\text{MBC}}(\omega_j)}{2\pi}}
  \left[ \oalphad(\omega)\op_j - \opd_j\oalpha(\omega)\right],
\end{equation}
where $\op_j$ is the annihilation (bosonic) operator
of a polariton in $j$-th eigen-transition-mode
and $\omega_j$ is its eigen-frequency.
Whereas we also get the explicit expression of loss rate $\kappa_{\text{MBC}}(\omega)$,
it is not easy to catch
whether $\oHSE^{\text{MBC}}$ is expressed by electric field [$\oX^- = \ii(\oa-\oad)$]
or the magnetic one [$\oX^+ = \oa + \oad$],
because Eq.~\eqref{eq:oHSE_MBC} corresponds to the SEC Hamiltonian
\eqref{eq:oHSE_pre} after the pre-trace RWA.
Further, the derivation of the SEC Hamiltonian is applicable to mainly the bosonic excitations in matters
($\op_j$ and $\omega_j$ are derived by Bogoliubov transformation),
and the extension to general nonlinear systems (e.g., Jaynes-Cummings and Rabi models with/without $A^2$ term)
is not straightforward.

In the following discussions, we try to derive the SEC Hamiltonian
from the Lagrangian describing the detailed mechanism of the SEC.

\section{Straightforward derivation} \label{sec:straight}
\begin{figure}[tbp]
\includegraphics[width=\linewidth]{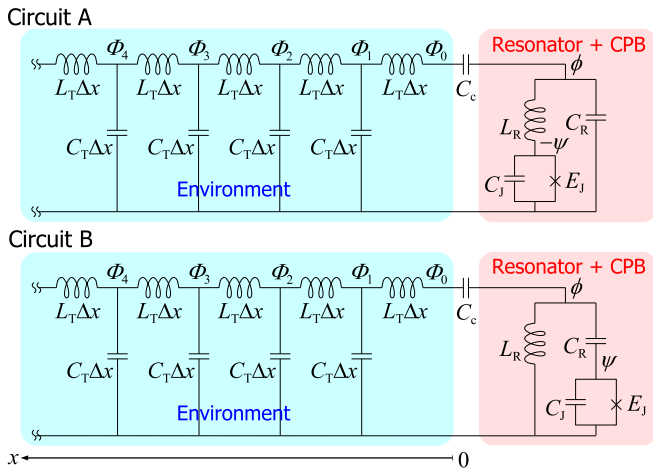}
\caption{In both circuits,
a LC-resonator with capacitance $\CR$ and inductance $\LR$
couples with a semi-infinite transmission line (environment)
with capacitance $\CT$ and inductance $\LT$ per unit length.
They are coupled by a capacitance $\CC$.
The resonator also couples with
a Cooper pair box (CPB) with capacitance $\CJ$ and Josephson energy $\EJ$.
Whereas the resonator-CPB coupled system itself is equivalent in both circuits
and the resonator-environment coupling is also equivalent,
the two circuits are not equivalent as a whole due to the difference
of the position of the CPB,
which corresponds to the difference of the inductive resonator-CPB interaction (circuit A)
and the capacitive one (circuit B).}
\label{fig:1}
\end{figure}
In this and the next sections,
we consider explicitly superconducting circuits,
which have a good correspondence with the cavity QED
and have a rich tunability compared with the actual light-matter system
\cite{Blais2004PRA,Devoret2007AP},
especially the circuits depicted in Fig.~\ref{fig:1}.
Before deriving Hamiltonians,
we must derive Lagrangians describing correctly the circuits,
and the derivation procedure of them is already established
\cite{Yurke1984PRA,Devoret1997}.

The Lagrangian applicable to both two circuits
is shown in Sec.~\ref{sec:Lagrangian_LC},
and we try to derive the Hamiltonian with remaining artificial atoms
(matters) as a black box.
Although the SEC Hamiltonian cannot be derived without the knowledge
inside the cavity in general,
we can derive it in the special case where
the resonator fields interact inductively with artificial atoms
as circuit A.
This situation is discussed in Sec.~\ref{sec:LCB-flux}.
In the opposite case, where
the resonator fields interact with artificial atoms capacitively
as circuit B,
the expression of the SEC Hamiltonian depends on
the detail of the cavity system.
In Sec.~\ref{sec:LCA-charge},
we derive the SEC Hamiltonian by considering
circuit B explicitly.
The difference of dissipative behaviors of the two circuits
is demonstrated in Sec.~\ref{sec:demo_straight}.

\subsection{Lagrangian} \label{sec:Lagrangian_LC} % !!!!!!!!!!!!!!!!!!!!!!!!!!
In both circuits in Fig.~\ref{fig:1},
a LC-resonator characterized by capacitance $\CR$ and inductance $\LR$
confines the superconducting current (microwave) with a frequency of $\wz = 1/\sqrt{\LR\CR}$,
which corresponds to the electromagnetic fields in a cavity \cite{Blais2004PRA}.
At the capacitance $\CC$,
the resonator couples with a semi-infinite transmission line (external line, environment)
characterized by capacitance $\CT$ and inductance $\LT$ per unit length.
The difference of circuits A and B is the positions of the Cooper pair box
(CPB, artificial atoms)
characterized by capacitance $\CJ$ and Josephson energy $\EJ$,
which couples with the resonator and corresponds to matters
interacting with the cavity fields.
The two circuits are equivalent in the absence of the CPB
or of the external line.
However, due to the presence of both, the two circuits are in principle different.
Then, the dissipative motion of the resonator fields depends on
whether the resonator-CPB interaction
are mediated by the charge (capacitive) or the magnetic flux (inductive) in the LC-resonator,
which corresponds to the electric and magnetic fields in a cavity of light.

Once a circuit is given,
we can derive its Lagrangian \cite{Yurke1984PRA,Devoret1997},
and then its Hamiltonian is also determined.
However, here we try to derive the SEC Hamiltonian
with remaining the matter system as a black box
(not only the CPB but also other elements can exist).
By defining the flux $\phi$ in the resonator
and $\{\varPhi_j\}$ in the external line as in Fig.~\ref{fig:1},
the Lagrangian of the whole circuit is generally expressed as
\cite{Devoret1997,Wallquist2006PRB,Johansson2010PRA}
\begin{equation} \label{eq:L_A-LC-TL} % !!!!!!!!!!!!!!!!!!!!!!!!!!!!!!!!!!!!!!
\LL = \Lz
+ \frac{\CC}{2}(\dot{\phi} - \dot{\varPhi}_0)^2
+ \sum_{j=1}^{\infty}\left[
    \frac{\CT\dx}{2}\dot{\varPhi}_j{}^2
  - \frac{(\varPhi_j-\varPhi_{j-1})^2}{2\LT\dx}
  \right].
\end{equation}
The last two terms are the Lagrangian of the external line.
$\dx$ is the infinitesimal distance,
and the dot ( $\dot{}$ ) means the time-derivative.
The second term represents the SEC,
and $\Lz$ is the Lagrangian in the resonator:
\begin{equation}
\Lz = \frac{\CR}{2}\dot{\phi}{}^2 - \frac{\phi{}^2}{2\LR} + \LA,
\end{equation}
where $\LA$ expresses arbitrary circuit elements such as the CPB
and their interactions with the resonator fields.
The conjugate momenta are obtained as follows:
\begin{subequations}
\begin{align}
q & = \frac{\partial \LL}{\partial \dot{\phi}}
    = \CR\dot{\phi} + \frac{\partial \LA}{\partial \dot{\phi}}
    + \CC(\dot{\phi} - \dot{\varPhi}_0),
\label{eq:q_LCB-TL} \\ % !!!!!!!!!!!!!!!!!!!!!!!!!!!!!!!!!!!!!!!!!!!!!!!!!!!!!
Q_0
& = \frac{\partial \LL}{\partial \dot{\varPhi}_0}
  = \CC(\dot{\varPhi}_0 - \dot{\phi}),
\label{eq:Q_LCB-TL} \\ % !!!!!!!!!!!!!!!!!!!!!!!!!!!!!!!!!!!!!!!!!!!!!!!!!!!!!
Q_j
& = \frac{\partial \LL}{\partial \dot{\varPhi}_j}
  = \CT\dx\dot{\varPhi}_j \quad \text{for $j > 0$}.
\end{align}
\end{subequations}
Then, the Lagrange equations are obtained as
\begin{subequations} \label{eq:Lagrange_LC-TL_dis} % !!!!!!!!!!!!!!!!!!!!!!!!!
\begin{align}
&\CR\ddot{\phi} + \frac{\dd}{\dd t}\frac{\partial \LA}{\partial \dot{\phi}} + \CC(\ddot{\phi} - \ddot{\varPhi}_0)
 = - \frac{\phi}{\LR} + \frac{\partial \LA}{\partial \phi}, \\
&\CC(\ddot{\varPhi}_0 - \ddot{\phi}) = - \frac{\varPhi_0 - \varPhi_1}{\LT\dx}, \\
&\CT\dx\ddot{\varPhi}_j = \frac{\varPhi_{j+1}+\varPhi_{j-1}-2\varPhi_j}{\LT\dx} \quad \text{for $j > 0$}.
\end{align}
\end{subequations}
In the continuous description $\varPhi(x_j) = \varPhi_j$ of the field in the transmission line,
these equations are rewritten as
\begin{subequations} \label{eq:Lagrange_LC-TL} % !!!!!!!!!!!!!!!!!!!!!!!!!!!!!
\begin{align}
& \CR\ddot{\phi} + \frac{\dd}{\dd t}\frac{\partial \LA}{\partial \dot{\phi}} + \CC[\ddot{\phi} - \ddot{\varPhi}(0^+)]
= - \frac{\phi}{\LR} + \frac{\partial \LA}{\partial \phi}, \label{eq:Lagrange_phi} \\
&\CC[\ddot{\varPhi}(0^+) - \ddot{\phi}] = \frac{1}{\LT}\left.\ddx{\varPhi(x)}\right|_{x=0^+}
\label{eq:boundary_LC-TL}, \\
&\CT\ddot{\varPhi}(x) = \frac{1}{\LT}\ddxx{\varPhi(x)} \quad \text{for $x > 0$}.
\label{eq:wave_LC-TL} % !!!!!!!!!!!!!!!!!!!!!!!!!!!!!!!!!!!!!!!!!!!!!!!!!!!!!!
\end{align}
\end{subequations}
As seen in Eq.~\eqref{eq:wave_LC-TL},
the fields in the external line
simply propagate with a speed of $v = 1 / \sqrt{\LT\CT}$,
while it contacts with the resonator at the capacitance $\CC$ ($x=0$)
and Eq.~\eqref{eq:boundary_LC-TL} is the boundary condition.
From Eqs.~\eqref{eq:Lagrange_phi} and \eqref{eq:boundary_LC-TL},
we get
\begin{equation}
\CR\ddot{\phi}  + \frac{\phi}{\LR} + \frac{\dd}{\dd t}\frac{\partial \LA}{\partial \dot{\phi}} - \frac{\partial \LA}{\partial \phi} = \frac{1}{\LT}\left.\ddx{\varPhi(x)}\right|_{x=0^+}.
\end{equation}
The left-hand side is the equation of motion
of $\phi$ derived from the Lagrangian $\Lz$ of the resonator system,
and the right-hand side is the perturbation due to the coupling
with the environment (external line).
We can fully solve the equations of motion with the knowledge of $\LA$.
But, can we derive the SEC Hamiltonian without it?
The total Hamiltonian is derived in terms of the fluxes and their time-derivatives as
\begin{align} \label{eq:H_LC-TL_derivative} % !!!!!!!!!!!!!!!!!!!!!!!!!!!!!
& \HH(\phi,\dot{\phi},\ldots,\{\varPhi_j,\dot{\varPhi}_j\})
= \Hz(\phi,\dot{\phi},\ldots)
+ \frac{\CC}{2}(\dot{\phi}-\dot{\varPhi}_0)^2 \nonumber \\ & \quad
+ \sum_{j=1}^{\infty}\left[
    \frac{\CT\dx}{2}\dot{\varPhi}_j{}^2
  + \frac{(\varPhi_j-\varPhi_{j-1})^2}{2\LT\dx}
  \right],
\end{align}
where $\Hz$ is the Hamiltonian derived from $\Lz$.
The SEC is represented by the second term.
In terms of the conjugate momenta,
the Hamiltonian is rewritten as
\begin{align} \label{eq:H_LC-TL} % !!!!!!!!!!!!!!!!!!!!!!!!!!!!!!!!!!!!!!!!
& \HH(\phi,q,\ldots,\{\varPhi_j,Q_j\})
= \Hz(\phi,q+Q_0,\ldots) \nonumber \\ & \quad
+ \frac{Q_0{}^2}{2\CC}
+ \sum_{j=1}^{\infty}\left[
    \frac{Q_j{}^2}{2\CT\dx}
  + \frac{(\varPhi_j-\varPhi_{j-1})^2}{2\LT\dx}
  \right].
\end{align}
In this way, when we consider the capacitive coupling between
the resonator and the external line
as in Fig.~\ref{fig:1}
(and employing the flux-base description \cite{Devoret1997}),
the SEC is included in $\oHz$ and determined
with the parameters of $\Hz$ replacing $q$ by $q+Q_0$.
Whereas the SEC Hamiltonian has been expressed simply by $\phi$ or $q$
in the well-known discussions \cite{gardiner04,Breuer2006,Beaudoin2011PRA},
it is derived in a complicated form in general,
e.g., expressed by $q/\CR+(q+\rho)/\CJ$
with the charge $\rho$ in the CPB for circuit B
as will be shown in Sec.~\ref{sec:LCA-charge}.
Then, the SEC Hamiltonian in principle
depends on the detail of the resonator system,
and we cannot clearly distinguish the Hamiltonians of it,
of environment, and of the SEC
without the knowledge of $\LA$.
However, we can do in specific situations such as circuit A in Fig.~\ref{fig:1}.
In the next subsection, we first derive the SEC Hamiltonian in such situations
by the straightforward calculation,
although it gives the complicated SEC Hamiltonian for circuit B
as discussed in Sec.~\ref{sec:LCA-charge}.

\subsection{In case of inductive resonator-CPB interaction (Circuit A)} \label{sec:LCB-flux}
As a less general case,
if the other elements in the LC-resonator
interact with the resonator fields
only by its flux $\phi$ but not by the time-derivative $\dot{\phi}$
(this means $\partial \LA/\partial \dot{\phi} = 0$),
the total Hamiltonian can be separated clearly as
$\HH^A = \Hz^A(\phi,q,\ldots) + \HSE^A + \HE^A$.
Here, $\Hz^A = \HLC + \ldots$ includes the Hamiltonian of the LC-resonator
\begin{equation}
\HLC = \frac{q^2}{2\CR} + \frac{\phi^2}{2\LR},
\end{equation}
and also represents arbitrary circuit elements (artifical atoms, other resonators, etc.)
and the interaction between them.
Even if the resonator embeds complicated elements,
if $\partial \LA/\partial \dot{\phi} = 0$ is satisfied,
the SEC Hamiltonian is represented simply as
\begin{equation} \label{eq:HSE_A} % !!!!!!!!!!!!!!!!!!!!!!!!!!!!!!!!!!!!!!!!!!
\HSE^A = \frac{qQ_0}{\CR},
\end{equation}
and the environment is defined as
\begin{equation}
\HE^A = \frac{Q_0{}^2}{2\CCR}
+ \sum_{j=1}^{\infty}\left[
    \frac{Q_j{}^2}{2\CT\dx}
  + \frac{(\varPhi_j-\varPhi_{j-1})^2}{2\LT\dx}
  \right],
\end{equation}
where the contact capacitance is modulated as
\begin{equation}
\frac{1}{\CCR} = \frac{1}{\CC} + \frac{1}{\CR}.
\end{equation}

The Hamiltonian of the LC-resonator can be easily quantized as
\begin{equation}
\oHLC = \hbar\wz\oad\oa,
\end{equation}
where the annihilation and creation operators satisfy
\begin{subequations}
\begin{align}
\left[ \oa, \oad \right] & = 1, \\
\left[ \oa, \oa \right] & = 0.
\end{align}
\end{subequations}
In terms of them, the resonator fields are expressed as
\begin{subequations} \label{eq:ophi_oq_a} % !!!!!!!!!!!!!!!!!!!!!!!!!!!!!!!!!!
\begin{align}
\ophi & = \sqrt{\frac{\hbar\ZR}{2}}(\oa + \oad), \\
\oq & = -\ii\sqrt{\frac{\hbar}{2\ZR}}(\oa-\oad),
\end{align}
\end{subequations}
where $\ZR = \sqrt{\LR/\CR}$ is the impedance of the LC-circuit.
They certainly satisfy $[ \ophi, \oq ] = \ii\hbar$.

We next quantize the environment $\HE$.
From this Hamiltonian,
the equation of motion of quantized external field $\oPhi(x)$ is derived
in the continuous basis as the wave equation for $x > 0$:
\begin{equation} \label{eq:wave-Phi} % !!!!!!!!!!!!!!!!!!!!!!!!!!!!!!!!!!!!!!!
\CT\ddtt{}\oPhi(x) = \frac{1}{\LT}\ddxx{\oPhi(x)} \quad \text{for $x > 0$}.
\end{equation}
When the environment $\HE$ is isolated from the resonator system,
the external fields obey the following boundary conditions
at the boundary $x = 0^+$:
\begin{subequations}
\begin{align}
\ddt{}\oPhi(0^+) & = \frac{\oQ_0}{\CCR} \label{eq:dPhi0=Q0/CCR}, \\ % !!!!!!!!!
\ddt{}\oQ_0 & = \frac{1}{\LT}\left.\ddx{\oPhi(x)}\right|_{x=0^+}.
\end{align}
\end{subequations}
From these two equations, we get the following relation for $\varPhi(x=0^+)$
\begin{equation}
\CCR\ddtt{}\oPhi(0^+) = \frac{1}{\LT}\left.\ddx{\oPhi(x)}\right|_{x=0^+}.
\end{equation}
Solving the wave equation \eqref{eq:wave-Phi} with this boundary condition,
the wave function at frequency $\omega$ is obtained as
\begin{equation}
f(z,\omega) = \sqrt{\frac{2}{\len}}\frac{\cos(\omega x/v) - \ii\modu(\omega)\sin(\omega x/v)}{1+\ii\modu(\omega)},
\end{equation}
where $\modu(\omega)$ is the non-dimensional parameter
defined with the characteristic impedance $\ZT = \sqrt{\LT/\CT}$ of the transmission line:
\begin{equation}
\modu(\omega) = \omega \ZT\CCR.
\end{equation}
Then, the positive-frequency components of the external fields
are quantized by annihilation operators $\{\oalpha_m\}$ as
\begin{subequations}
\begin{align}
\oPhi^+(z)
& = \sum_{m=1}^{\infty} \sqrt{\frac{\hbar\ZT}{k_m\len}}
    \frac{\cos(k_mx) - \ii\modu(vk_m)\sin(k_mx)}{1+\ii\modu(vk_m)}
    \oalpha_m, \label{eq:oPhi+(z)} \\ % !!!!!!!!!!!!!!!!!!!!!!!!!!!!!!!!!!!!!!!
\oQ^+(z)
& = - \ii \sum_{m=1}^{\infty} \sqrt{\frac{\hbar k_m}{\ZT\len}}
    \frac{\cos(k_mx) - \ii\modu(vk_m)\sin(k_mx)}{1+\ii\modu(vk_m)}
    \oalpha_m,
\end{align}
\end{subequations}
where the wavenumber is defined as $k_m = m\pi / \len$
with length $\len$ of the external system
and integer $m = 1, 2, \ldots$.
The Hamiltonian of the environment is rewritten as
\begin{equation}
\oHE^A
= \sum_{m=1}^{\infty} \hbar v k_m \oalphad_m\oalpha_m.
\end{equation}

From Eq.~\eqref{eq:dPhi0=Q0/CCR}, $\oQ_0$ is described by the external field as
\begin{subequations}
\begin{align}
\oQ_0^+
& = - \ii v \CCR \sum_{m=1}^{\infty} \sqrt{\frac{\hbar k_m\ZT}{\len}}
    \frac{\oalpha_m}{1+\ii\modu(vk_m)}, \\
& = - \ii \CCR \int_0^{\infty}\dd\omega\
    \frac{\sqrt{2\hbar\omega\ZT}}{1+\ii\modu(\omega)}
    \frac{\oalpha(\omega)}{\sqrt{2\pi}}.
\end{align}
\end{subequations}
Therefore, the SEC Hamiltonian \eqref{eq:HSE_A} is expressed as
\begin{equation} \label{eq:oHSE-LCB-flux} % !!!!!!!!!!!!!!!!!!!!!!!!!!!!!!!!!!
\oHSE^A
= \int_0^{\infty}\dd\omega\
  \ii\hbar\sqrt{\frac{\kappaLCz'}{2\pi}\frac{\omega}{\wz}
    \frac{2\ZR}{\hbar}} \oq
  \left[ \frac{\oalphad(\omega)}{1-\ii\modu(\omega)}
    - \Hc \right],
\end{equation}
where $\kappaLCz'$ is the loss rate of bare resonator mode
\begin{equation}
\kappaLCz'
= \frac{\wz\ZT\CCR{}^2}{\ZR\CR{}^2}.
\end{equation}
In this way, when the interaction between LC-resonator and other elements
is inductive ($\partial \LA/\partial\dot{\phi} = 0$)
as in circuit A, the SEC Hamiltonian
can be simply expressed by $\oq$ as Eq.~\eqref{eq:oHSE-LCB-flux}
not by $\ophi$.
Then, when we define the annihilation operator $\oa$ of the resonator fields
as Eqs.~\eqref{eq:ophi_oq_a},
this SEC Hamiltonian \eqref{eq:oHSE-LCB-flux}
corresponds to $\oHSE^-$ \eqref{eq:oHSE_-} as
\begin{equation} \label{eq:oHSE_A_a} % !!!!!!!!!!!!!!!!!!!!!!!!!!!!!!!!!!!!!!!
\oHSE^A = \int_0^{\infty}\dd\omega\
  \hbar\sqrt{\frac{\kappabare^A(\omega)}{2\pi}}(\oa-\oad)
  \left[ \oalphad(\omega) - \Hc \right],
\end{equation}
where the frequency-dependent bare loss rate is
\begin{equation}
\kappabare^{A}(\omega)
= \frac{\kappaLCz'}{|1-\ii\zeta(\omega)|^2}\left(\frac{\omega}{\wz}\right).
\end{equation}
In the next subsection, we consider a more general situation where $\partial \LA/\partial \dot{\phi} \neq 0$.
In this case, the SEC Hamiltonian is modified
by the detail of the resonator system,
and then we consider circuit B explicitly.
We will find that the SEC Hamiltonian is expressed
not simply as Eq.~\eqref{eq:oHSE-LCB-flux} in general.

\subsection{In case of capacitive resonator-CPB interaction (Circuit B)} \label{sec:LCA-charge}
Whereas the SEC Hamiltonian cannot be derived in the straightforward way
in general situation $\partial \LA/\partial \dot{\phi} \neq 0$,
it is possible derived when we consider a circuit explicitly, such as circuit B in Fig.~\ref{fig:1}.
When we define $\psi$ as the flux through the CPB as shown in Fig.~\ref{fig:1},
the Lagrangian of the resonator system is represented as
\begin{equation} \label{eq:Lz_B} % !!!!!!!!!!!!!!!!!!!!!!!!!!!!!!!!!!!!!!!!!!!
\Lz^B = \frac{\CR}{2}(\dot{\phi}-\dot{\psi})^2
- \frac{\phi^2}{2\LR} + \frac{\CJ}{2}\dot{\psi}^2 + \EJ\cos(2e\psi/\hbar).
\end{equation}
Then, the total Hamiltonian including the external line is obtained as follows
\begin{equation}
\HH^B = \Hz^B + \HSE^B + \frac{Q_0{}^2}{2\CCCR}
  + \sum_{j=1}^{\infty}\left[
    \frac{Q_j{}^2}{2\CT\dx}
  + \frac{(\varPhi_j-\varPhi_{j-1})^2}{2\LT\dx}
  \right],
\end{equation}
where the contact capacitance is modulated in this case as
\begin{equation}
\frac{1}{\CCCR} = \frac{1}{\CC} + \frac{1}{\CR} + \frac{1}{\CJ}.
\end{equation}
The Hamiltonian $\Hz^B$ is the one derived from $\Lz^B$:
\begin{equation} \label{eq:Hz_B} % !!!!!!!!!!!!!!!!!!!!!!!!!!!!!!!!!!!!!!!!!!!
\Hz^B = \frac{q^2}{2\CR} + \frac{\phi^2}{2\LR} 
+ \frac{(q+\rho)^2}{2\CJ} - \EJ\cos(2e\psi/\hbar),
\end{equation}
where $\rho = \partial\Lz^B/\partial \dot{\psi}$ is the conjugate momentum of $\psi$.
The SEC Hamiltonian is derived as
\begin{equation} \label{eq:HSE_B} % !!!!!!!!!!!!!!!!!!!!!!!!!!!!!!!!!!!!!!!!!!
\HSE^B = \left( \frac{q}{\CR} + \frac{q+\rho}{\CJ} \right)Q_0.
\end{equation}
In the same manner as the previous subsection,
the quantized external field at capacitance $\CC$ is represented as
\begin{equation}
\oQ_0^+ = - \ii \CCCR \int_0^{\infty}\dd\omega\ \frac{\sqrt{2\hbar\omega\ZT}}{1+\ii\zeta'(\omega)} \frac{\oalpha(\omega)}{\sqrt{2\pi}},
\end{equation}
where $\zeta'(\omega) = \omega\ZT\CCCR$.
Then, the SEC Hamiltonian is expressed as
\begin{align} \label{eq:oHSE-A} % !!!!!!!!!!!!!!!!!!!!!!!!!!!!!!!!!!!!!!!!!
\oHSE^B
& = \int_0^{\infty}\dd\omega\ \ii\CCCR \sqrt{\frac{2\hbar\omega\ZT}{2\pi}}
\nonumber \\ & \quad \times
  \left(\frac{\oq}{\CR} + \frac{\oq+\orho}{\CJ}\right)
  \left[ \frac{\oalphad(\omega)}{1-\ii\zeta'(\omega)} - \Hc \right].
\end{align}
In this way, by the straightforward calculation,
we get such a complicated expression as 
the SEC Hamiltonian compared with Eq.~\eqref{eq:oHSE-LCB-flux},
which is expressed simply by $\oq$.
This expression cannot be reduced to the simple ones as Eqs.~\eqref{eq:oHSE_pm},
which we usually introduce intuitively.
Further, Eq.~\eqref{eq:oHSE-A} is the expression specially for circuit B,
and the SEC Hamiltonian is modified depending on the detail of $\LA$ in general (when $\partial \LA/\partial \dot{\phi} \neq 0$).
However, as will be shown in Sec.~\ref{sec:derive_SEC},
in the good-cavity and independent-transition limit,
we can derive the SEC Hamiltonian in a simple form as Eqs.~\eqref{eq:oHSE_pm},
and its expression is independent of $\LA$ even if $\partial \LA/\partial \dot{\phi} \neq 0$.

\subsection{Demonstration} \label{sec:demo_straight}
Before deriving the general expression of the SEC,
here we demonstrate the difference of the dissipative behaviors
in circuits A and B of Fig.~\ref{fig:1}.
Whereas the Lagrangian $\Lz^B$ for circuit B is derived in Eq.~\eqref{eq:Lz_B},
the one for circuit A is obtained as
\begin{equation}
\Lz^A = \frac{\CR}{2}\dot{\phi}^2 - \frac{(\phi+\psi)^2}{2\LR} + \frac{\CJ}{2}\dot{\psi}^2 + \EJ\cos(2e\psi/\hbar).
\end{equation}
In the absence of the external line, the two circuits are equivalent.
The difference between $\Lz^A$ and $\Lz^B$ comes from the definition
of $\psi$ and the position of the earth for the two circuits as drawn in Fig.~\ref{fig:1}
\cite{Devoret1997}.
From Lagrangian $\Lz^A$, we get the corresponding Hamiltonian as
\begin{equation} \label{eq:Hz_A} % !!!!!!!!!!!!!!!!!!!!!!!!!!!!!!!!!!!!!!!!!!!
\Hz^A = \frac{q^2}{2\CR} + \frac{(\phi+\psi)^2}{2\LR}
+ \frac{\rho^2}{2\CJ} - \EJ\cos(2e\psi/\hbar).
\end{equation}
Here, $\rho = \partial\Lz^A/\partial\dot{\psi}$ is the conjugate momentum of $\psi$.
The expressions of $\Hz^A$ and $\Hz^B$ in Eq.~\eqref{eq:Hz_B} looks different,
and the interactions between the LC-resonator and the CPB
seem inductive (through $\phi$ and $\psi$)
and capacitive (through $q$ and $\rho$) for circuits A and B, respectively.
However, they are related as
\begin{equation} \label{eq:HzA=UHzBU} % !!!!!!!!!!!!!!!!!!!!!!!!!!!!!!!!!!!!!!
\Hz^A = U^* \Hz^B U,
\end{equation}
where $U$ is the unitary operator defined as
\begin{equation}
U = \exp(-\ii q \psi/\hbar).
\end{equation}
By using this, we also get
\begin{subequations}
\begin{align}
U^* \phi U & = \phi + \psi, \\
U^* q U & = q, \\
U^* \psi U & = \psi, \\
U^* \rho U & = \rho - q.
\end{align}
\end{subequations}
Then, $\Hz^A$ and $\Hz^B$ are in principle equivalent,
and give the same eigen-frequencies,
because circuits A and B are equivalent in the absence of the external line.
Then, we cannot distinguish the inductive and capacitive interactions
in such simple circuits without the external line.
Depending on whether the interaction is capacitive or inductive
\cite{Niemczyk2010NP,Fedorov2010PRL,Forn-Diaz2010PRL}
in real circuits of interest,
the position of the CPB is modeled as Fig.~\ref{fig:1}.
However, in the presence of the external line,
the two circuits are clearly different,
and the Hamiltonians are also not equal as a whole.
Even by transforming the SEC Hamiltonian $\HSE^B$ \eqref{eq:HSE_B}
of circuit B as
\begin{equation} \label{eq:Ud_HSEB_U} % !!!!!!!!!!!!!!!!!!!!!!!!!!!!!!!!!!!!!!
U^*\HSE^BU
= \left( \frac{q}{\CR} + \frac{\rho}{\CJ} \right)Q_0,
\end{equation}
this is not equivalent with $\HSE^A$ \eqref{eq:HSE_A} of circuit A.

For simplicity, we neglect the anharmonicity in the CPB
as $\cos(2e\opsi/\hbar) \sim 1 + (2e/\hbar)^2\opsi^2$
(in the transmon regime \cite{Devoret2007AP,Bourassa2012PRA}
but general anharmonic systems can be considered
as discussed in Sec.~\ref{sec:review}),
and then the fields relevant to the CPB can be expressed by bosonic operator
$\ob$ satisfying
\begin{subequations}
\begin{align}
\left[ \ob, \obd \right] & = 1, \\
\left[ \ob, \ob \right] & = 0.
\end{align}
\end{subequations}
The Hamiltonian of the CPB is rewritten as
\begin{equation}
\frac{\orho^2}{2\CJ} - \EJ\cos(2e\opsi/\hbar)
\simeq \hbar\wx\obd\ob + \const,
\end{equation}
and the fields are represented as
\begin{subequations}
\begin{align}
\opsi & = \frac{\hbar}{2e}\left(\frac{\ECP}{2\EJ}\right)^{1/4} (\ob+\obd), \\
\orho & = -\ii\frac{\hbar\wx\CJ}{2e}\left(\frac{\ECP}{2\EJ}\right)^{1/4} (\ob-\obd)
\nonumber \\ &
= -\ii\frac{2e}{2}\left(\frac{2\EJ}{\ECP}\right)^{1/4} (\ob-\obd).
\end{align}
\end{subequations}
Here, $\wx = \sqrt{2\ECP\EJ}/\hbar$ is the excitation frequency
and $\ECP = (2e)^2/2\CJ$ is the capacitive energy in the CPB \cite{Devoret2007AP}.
In terms of the annihilation and creation operators,
the Hamiltonians of resonator systems are represented as
\begin{subequations} \label{eq:oHz-L-C} % !!!!!!!!!!!!!!!!!!!!!!!!!!!!!!!!!!!!
\begin{align}
\oHz^{A} & = \hbar\wz\oad\oa + \hbar\wx\obd\ob
\nonumber \\ & \quad
+ \hbar\rabi_A(\oa+\oad)(\ob+\obd)
+ \frac{\hbar\rabi_A{}^2}{\wz}(\ob+\obd)^2.
\label{eq:Hz-L-L} \\ % !!!!!!!!!!!!!!!!!!!!!!!!!!!!!!!!!!!!!!!!!!!!!!!!!!!!!!!
\oHz^{B} & = \hbar\wz\oad\oa + \hbar\wx\obd\ob
\nonumber \\ & \quad
- \hbar\rabi_B(\oa-\oad)(\ob-\obd)
+ \frac{\hbar\rabi_B{}^2}{\wx}[\ii(\oa-\oad)]^2,
\label{eq:Hz-C-L} % !!!!!!!!!!!!!!!!!!!!!!!!!!!!!!!!!!!!!!!!!!!!!!!!!!!!!!!
\end{align}
\end{subequations}
The interaction strength is defined as
\begin{equation}
\grabi
= \frac{\rabi_A}{\wz}
= \frac{\rabi_B}{\wx}
= \frac{1}{2e}\left(\frac{\ECP}{2\EJ}\right)^{1/4}\sqrt{\frac{\hbar}{2\ZR}},
\end{equation}
and this can reach to the ultrastrong regime $\grabi \sim 0.1$
\cite{Devoret2007AP},
while the anharmonicity might be small \cite{Bourassa2012PRA}.
The two Hamiltonians \eqref{eq:oHz-L-C} are also related
by the unitary operator
\begin{equation}
\oU = \exp[-\grabi(\oa-\oad)(\ob+\obd)]
\end{equation}
as $\oHz^A = \oUd\oHz^B\oU$
\cite{Devoret1997,Todorov2012PRB}.
However, in the presence of the external line,
the two circuits are different as a whole,
and then the fields in them are dissipated differently in general.

The two Hamiltonians \eqref{eq:oHz-L-C} can be diagonalized by the Bogoliubov transformation.
We introduce the annihilation operator of a polariton
(eigen-mode; superposition of resonator field and CPB one)
for circuit $\xi = A, B$ as \cite{Ciuti2005PRB,Bamba2013MBC}
\begin{equation}
\op_j^{\xi} = w_j^{\xi} \oa + x_j^{\xi} \ob + y_j^{\xi} \oad + z_j^{\xi} \obd.
\end{equation}
The coefficients are determined for satisfying
\begin{align}
\left[ \op_j^{\xi}, \oHz^{\xi}\right] & = \hbar\omega_j \op_j^{\xi}, \\
\left[ \op_j^{\xi}, \op^{\xi\dagger}_{j'} \right] & = \delta_{j,j'}.
\end{align}
Whereas the coefficients are obtained differently for the two circuits,
the eigen-frequencies $\omega_j$ of lower and upper polaritons ($j = L$ and $U$)
are obtained equally.
In terms of the polariton operators, the original ones are presented as
\begin{subequations}
\begin{align}
\oa & = \sum_{j=L,U}\left( w_j^{\xi*}\op_j^{\xi} - y_j^{\xi} \op_j^{\xi\dagger} \right), \\
\ob & = \sum_{j=L,U}\left( x_j^{\xi*}\op_j^{\xi} - z_j^{\xi} \op_j^{\xi\dagger} \right),
\end{align}
\end{subequations}
and then we also get
\begin{subequations}
\begin{align}
\oa + \oad & = \sum_{j=L,U}\left( w_j^{\xi} - y_j^{\xi} \right)^* \op_j^{\xi} + \Hc, \\
\ii(\oa - \oad) & = \sum_{j=L,U}\ii\left( w_j^{\xi} + y_j^{\xi} \right)^* \op_j^{\xi} + \Hc, \\
\ob + \obd & = \sum_{j=L,U}\left( x_j^{\xi} - z_j^{\xi}\right)^* \op_j^{\xi} + \Hc, \\
\ii(\ob - \obd) & = \sum_{j=L,U}\ii\left( x_j^{\xi} + z_j^{\xi} \right)^* \op_j^{\xi} + \Hc.
\end{align}
\end{subequations}
When we suppose the good-cavity limit $\CC \ll \CR$ for simplicity,
we get $\zeta(\omega) \ll 1$ and $\CCR \simeq \CCCR \simeq \CC$.
Performing the pre-trace RWA,
the SEC Hamiltonian $\oHSE^A$ \eqref{eq:oHSE_A_a} for circuit A is rewritten as
\begin{subequations} \label{eq:oHSE_A_p} % !!!!!!!!!!!!!!!!!!!!!!!!!!!!!!!!!!!
\begin{align}
\oHSE^A
& \simeq \int_0^{\infty}\dd\omega\
  \hbar\sqrt{\frac{\kappaLCz}{2\pi}\frac{\omega}{\wz}}(\oa-\oad)
  \left[ \oalphad(\omega) - \Hc \right], \\
& \simeq \int_0^{\infty}\dd\omega\
  \hbar\sqrt{\frac{\kappaLCz}{2\pi}\frac{\omega}{\wz}}\oalphad(\omega) 
  \sum_{j=L,U}\left( w_j^{A} + y_j^{A} \right)^* \op_j^{A}
\nonumber \\ & \quad
 + \Hc,
\end{align}
\end{subequations}
where
\begin{equation}
\kappaLCz
= \frac{\wz\ZT\CC{}^2}{\ZR\CR{}^2}.
\end{equation}
Then, the loss rates of the eigen-modes in circuit A are finally expressed as
\begin{equation} \label{eq:kappaA_q} % !!!!!!!!!!!!!!!!!!!!!!!!!!!!!!!!!!!!!!!
\kappa^{A}_j
= \kappaLCz\left(\frac{\omega}{\wz}\right)
  \left| w_j^{A} + y_j^{A} \right|^2
\end{equation}
For circuit B, the SEC Hamiltonian is obtained as Eq.~\eqref{eq:oHSE-A},
and it is rewritten in the good-cavity limit as
\begin{widetext}
\begin{subequations}
\begin{align}
\oHSE^B
& \simeq \int_0^{\infty}\dd\omega\ \hbar\sqrt{\frac{\kappaLCz}{2\pi}\frac{\omega}{\wz}}
  \left[
    \left(1+4\grabi{}^2\frac{\wx}{\wz}\right)(\oa-\oad)
  + 2\grabi\frac{\wx}{\wz}(\ob-\obd)
  \right]
  \left[ \oalphad(\omega) - \oalpha(\omega) \right] \\
& \simeq \int_0^{\infty}\dd\omega\ \hbar\sqrt{\frac{\kappaLCz}{2\pi}\frac{\omega}{\wz}}
  \oalphad(\omega) \sum_{j=L,U} \left[
    \left(1+4\grabi{}^2\frac{\wx}{\wz}\right)\left( w_j^{B} + y_j^{B} \right)
  + 2\grabi\frac{\wx}{\wz}\left( x_j^{B} + z_j^{B} \right) \right]^* \op_j^B + \Hc
\end{align}
\end{subequations}
Then, the loss rates of the eigen-modes in circuit B are obtained as
\begin{equation} \label{eq:kappaB_q} % !!!!!!!!!!!!!!!!!!!!!!!!!!!!!!!!!!!!!!!
\kappa^B_j = \kappaLCz\left(\frac{\omega}{\wz}\right)
\left|
    \left(1+4\grabi{}^2\frac{\wx}{\wz}\right)\left( w_j^{B} + y_j^{B} \right)
  + 2\grabi\frac{\wx}{\wz}\left( x_j^{B} + z_j^{B} \right) \right|^2.
\end{equation}
If we use the same Hamiltonian $\oHz^A = \oUd\oHz^B\oU$ of the resonator system
for both circuits,
the SEC Hamiltonian \eqref{eq:Ud_HSEB_U} for circuit B is represented as
\begin{subequations} \label{eq:oHSE_B_p} % !!!!!!!!!!!!!!!!!!!!!!!!!!!!!!!!!!!
\begin{align}
\oUd\oHSE^B\oU
& \simeq \int_0^{\infty}\dd\omega\ \hbar\sqrt{\frac{\kappaLCz}{2\pi}\frac{\omega}{\wz}}
  \left[ (\oa-\oad) + 2\grabi\frac{\wx}{\wz}(\ob-\obd)
  \right]
  \left[ \oalphad(\omega) - \oalpha(\omega) \right] \\
& \simeq \int_0^{\infty}\dd\omega\ \hbar\sqrt{\frac{\kappaLCz}{2\pi}\frac{\omega}{\wz}}
  \oalphad(\omega) \left[
    \left( w_j^{A} + y_j^{A} \right)
  + 2\grabi\frac{\wx}{\wz}\left( x_j^{A} + z_j^{A} \right) \right]^* \op_j^A + \Hc
\end{align}
\end{subequations}
\end{widetext}
Then, the loss rates in circuit B can also be expressed as
\begin{equation} \label{eq:kappaB_q2} % !!!!!!!!!!!!!!!!!!!!!!!!!!!!!!!!!!!!!!!
\kappa^B_j = \kappaLCz\left(\frac{\omega}{\wz}\right)
\left|
    \left( w_j^{A} + y_j^{A} \right)
  + 2\grabi\frac{\wx}{\wz}\left( x_j^{A} + z_j^{A} \right) \right|^2.
\end{equation}

\begin{figure}[tbp]
\begin{center}
\includegraphics[width=\linewidth]{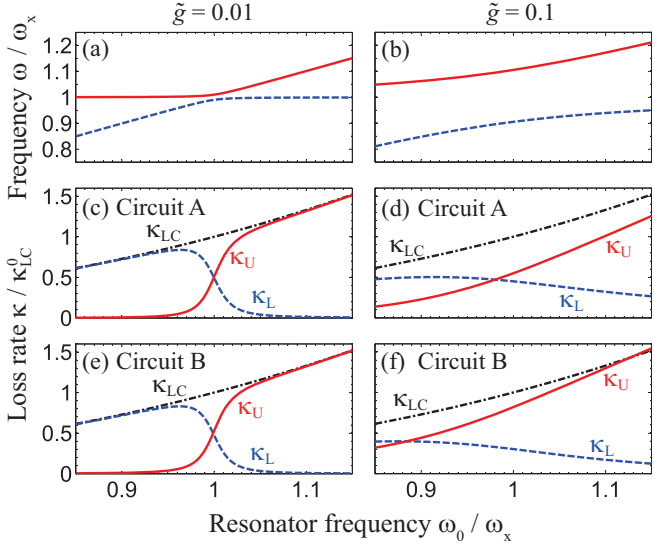}
\end{center}
\caption{The resonator-CPB interaction strength is supposed as
$\grabi = 0.01$ for (a,c,e) and $\grabi = 0.1$ for (b,d,f).
(a,b) The eigen-frequencies of lower and upper modes
(solid and dashed curves) in the resonator-CPB system of circuits A and B
are plotted versus the bare resonator frequency $\wz$
(the results are equivalent for the two circuits).
(c,d) The loss rates $\kappa_{L/U}^{A}$ of the two modes are plotted
for circuit A in Fig~\ref{fig:1}.
The dash-dotted curve represents $\kappaLCz$ of the bare resonator mode.
(e,f) The loss rates $\kappa_{L/U}^B$ for circuit B.
The loss rates are normalized to $\kappaLCz^0$ at $\wz = \wx$.
Although the two circuits have almost the same loss rates for $\grabi = 0.01$,
they are significantly different in the ultrastrong interaction regime
$\grabi = 0.1$.}
\label{fig:2}
\end{figure}
Here, we calculate the loss rates $\kappa^{\xi}_j$
\eqref{eq:kappaA_q}, \eqref{eq:kappaB_q}, and \eqref{eq:kappaB_q2}
for circuit $\xi = A, B$ and mode $j = L, U$.
By fabricating a SQUID at the other edge of the resonator,
we can effectively tune the resonator length by applying a magnetic flux
at the SQUID \cite{Johansson2009PRL,Johansson2010PRA,Wilson2011N}
($\CR$ and $\LR$ in Fig.~\ref{fig:1} are proportional to the effective resonator length),
and then $\wz$ can be tuned with keeping $\ZR$ and $\grabi$,
while the loss rate $\kappaLCz$ of bare resonator mode is modified.
In Fig.~\ref{fig:2}(a),
we plot the eigen-frequencies of the Hamiltonians \eqref{eq:oHz-L-C}
by changing the resonator frequency $\wz$ for fixed $\grabi = 0.01$.
$\oHz^{A}$ and $\oHz^B$ give the same eigen-frequencies:
the lower and upper modes (solid and dashed curves).
In Figs.~\ref{fig:2}(c) and (e),
we plot the loss rates $\kappaLC_{j}^{\xi}$ calculated
from the coefficients $\{w_j^{\xi}, x_j^{\xi}, y_j^{\xi}, z_j^{\xi}\}$
obtained by diagonalizing $\oHz^{A}$ and $\oHz^{B}$.
The dash-dotted curves represent the bare loss rate $\kappaLCz$,
and all the curves are normalized to $\kappaLCz^0$ at $\wz = \wx$.
We cannot find a clear difference between the two graphs
for such non-ultrastrong interaction $\grabi = 0.01$,
and $\kappaLC_{L/U} = \kappaLCz / 2$ is obtained at $\wz = \wx$
as in the standard discussion in quantum optics \cite{walls08}.

% Although the curves in Fig.~\ref{fig:2}(c) and (e) have a scalability for $\grabi \ll \wx$,
Whereas the loss rates of the two circuits are always equivalent in the case of $\grabi \ll 1$,
it is not the case in the ultrastrong interaction regime $\grabi = 0.1$
as seen in Figs.~\ref{fig:2}(d) and (f).
Whereas $\oHz^{A}$ and $\oHz^B$ still shows the same eigen-frequencies
for the upper and lower modes as in Fig.~\ref{fig:2}(b),
we get $\kappaLC_{j}^{\xi} \neq \kappaLCz / 2$ at $\wz = \wx$,
and the cross point $\kappaLC_L^{\xi} = \kappaLC_U^{\xi}$
is shifted to the left side (and the shifts are different for the two circuits).
In this way, the loss rates are strongly modulated in the ultrastrong interaction regime
$\grabi \gtrsim 0.1$,
and the dissipation of the resonator fields are clearly different
depending on whether the resonator-CPB interaction is inductive (circuit A)
or capacitive (circuit B)
even if the SEC is capacitive in both circuits.
When we consider the same Hamiltonian $\Hz^A = U^*\Hz^BU$
of the resonator system,
the loss rates in circuit B are given by Eq.~\eqref{eq:kappaB_q2},
and it gives the same values as Eq.~\eqref{eq:kappaB_q}
gives the same values for the loss rates in circuit B.
Whereas they are different from the loss rates given by Eq.~\eqref{eq:kappaA_q}
for circuit A, the difference is negligible
in the weak and normally strong light-matter interaction regimes $\grabi \ll 1$.

\section{General expression of SEC Hamiltonian} \label{sec:derive_SEC}
As discussed in the previous section,
even if the resonator system is equivalent as $\oHz^A = \oUd\oHz^B\oU$,
the SEC Hamiltonians are derived differently for circuits A and B
as Eqs.~\eqref{eq:oHSE_A_p} and \eqref{eq:oHSE_B_p}, respectively,
by the straightforward calculation.
Although the difference is negligible in the weak and normally strong
light-matter interaction regimes as seen in Fig.~\ref{fig:2},
in order to introduce correctly the dissipation of system-of-interest
in the ultrastrong light-matter interaction regime,
the SEC Hamiltonians must be derived
from the complete knowledge of systems-of-interest
and how they couple with environment.
This fact is not a good news for the theoretical studies
of ultrastrong light-matter interaction regime,
while we get another degree of freedom to control the dissipation and excitation.
However, as will be discussed in this section, 
when the quality of the cavity is good enough
and the transitions are independent,
we can derive a general expression of the SEC Hamiltonian
that is independent of the detail of system-of-interest,
even in the ultrastrong light-matter interaction regime.
In Sec.~\ref{sec:SEC_general}, the SEC Hamiltonian is derived
from the Lagrangian \eqref{eq:L_A-LC-TL} without specifying $\LA$.
The comparison with the straightforward ones is performed
in Sec.~\ref{sec:compare_general},
and the validity of the general expression is demonstrated
in Sec.~\ref{sec:demo_general}.

\subsection{Derivation} \label{sec:SEC_general}
In the good-cavity limit ($\kappaLCz \ll \wz, \wx$),
the time derivative of the flux $\phi$ in the LC-circuit
is approximately represented in terms of the lowering and raising operator $\osigma = \ket{\mu}\bra{\nu}$ as
\begin{equation}
\ddt{}\ophi \simeq \sum_{\mu,\nu}(-\ii\omega_{\nu,\mu})\braket{\mu|\ophi|\nu}\osigma_{\mu,\nu},
\end{equation}
where $\omega_{\nu,\mu} = \omega_{\nu} - \omega_{\mu}$
is the difference of the eigen-frequencies.
Then, since the SEC is expressed as $(\CC/2)(\dot{\phi}-\dot{\varPhi}_0)^2$
in the Lagrangian \eqref{eq:L_A-LC-TL} and also in the Hamiltonian \eqref{eq:H_LC-TL_derivative},
the equation of motion of $\osigma_{\mu,\nu}$ is approximately written as
\begin{subequations}
\begin{align}
\frac{\dd}{\dd t}\osigma_{\mu,\nu}
& = - \ii\omega_{\nu,\mu}\osigma_{\mu,\nu} + \frac{\CC}{2}\frac{1}{\ii\hbar}
    \left[ \osigma_{\mu,\nu}, (\dot{\phi}-\dot{\varPhi}_0)^2 \right], \\
& \simeq - \ii\omega_{\nu,\mu}\osigma_{\mu,\nu}
  - \sum_{\mu',\nu'} \frac{\omega_{\nu',\mu'}\CC}{\hbar}
    \braket{\nu'|\ophi|\mu'}
\nonumber \\ & \quad \times
    \left[ \osigma_{\mu,\nu}, \osigma_{\mu',\nu'} \right]
    \frac{\dd}{\dd t}(\ophi-\oPhi_0).
\end{align}
\end{subequations}
The last terms are the perturbation by the SEC.
Here, we also suppose that the transition $\{\mu,\nu\}$
of interest is well isolated from the others
(independent-transition limit).
Under this assumption, the above equation is approximately rewritten as
\begin{equation} \label{eq:motion_sigma} % !!!!!!!!!!!!!!!!!!!!!!!!!!!!!!!!!!!!!!!!!!!!!!!!!!!!
\frac{\dd}{\dd t}\osigma_{\mu,\nu}
\simeq - \ii\omega_{\nu,\mu}\osigma_{\mu,\nu}
  - \frac{\omega_{\nu,\mu}\CC}{\hbar}
    \left[ \osigma_{\mu,\nu}, \ophi \right]
    \frac{\dd}{\dd t}(\ophi-\oPhi_0).
\end{equation}
Here, we define the positive- and negative-frequency components
of the resonator and external fields as
\begin{subequations}
\begin{align}
\ophi(t) & = \int_0^{\infty}\dd\omega \left[ \ee^{-\ii\omega t} \ophi^+(\omega) + \ee^{\ii\omega t} \ophi^-(\omega) \right], \\
\oPhi(x,t) & = \int_0^{\infty}\dd\omega \left[ \ee^{-\ii\omega t} \oPhi^+(x,\omega) + \ee^{\ii\omega t} \oPhi^-(x,\omega) \right].
\end{align}
\end{subequations}
The positive-frequency components corresponds to the annihilation (lowering)
operator in the basis of the true eigen-states.
Namely, as derived in Eq.~\eqref{eq:oPhi+(z)} for circuit A,
in the good-cavity limit ($\zeta(\omega) \ll 1$),
the field $\oPhi^+(x)$ in the external line is described 
by annihilation operators $\oalpha(\omega)$ as
\begin{equation}
\oPhi^+(x,\omega) = \sqrt{\frac{\hbar\ZT}{\pi\omega}}
\cos\left(\omega x/v\right) \oalpha(\omega).
\end{equation}
On the other hand, $\ophi^+$ is expressed by the lowering
operators as
\begin{equation}
\ophi^+ = \sum_{\mu,\nu>\mu} \braket{\mu|\ophi|\nu} \osigma_{\mu,\nu}.
\end{equation}
Here, when the environment is large enough
and the quality of the resonator is good enough,
the negative-frequency components are negligible 
in the equation of motion of lowering operator
(corresponding to the pre-trace RWA in Sec.~\ref{sec:review}).
Then, Eq.~\eqref{eq:motion_sigma} is rewritten for $\nu > \mu$ as
\begin{align} \label{eq:motion_sigma_w_app} % !!!!!!!!!!!!!!!!!!!!!!!!!!!!!
\frac{\dd}{\dd t}\osigma_{\mu,\nu}
& \simeq - \ii\omega_{\nu,\mu}\osigma_{\mu,\nu}
  + \left[ \osigma_{\mu,\nu}, \ophi \right]
\nonumber \\ & \quad \times
    \int_0^{\infty}\dd\omega\
      \ee^{-\ii\omega t}
      \frac{\ii\omega\omega_{\nu,\mu}\CC}{\hbar}
      \left[\ophi^+(\omega)-\oPhi_0^+(\omega)\right].
\end{align}
On the other hand, since the dynamics in the external line (for $x > 0$)
is described by the simple wave equation \eqref{eq:wave_LC-TL},
the field $\oPhi^+(x)$ can be divided into the incoming field $\oPhiin^+$
and the outgoing one $\oPhiout^+$ as \cite{Bamba2013MBC}
\begin{equation}
\oPhi^+(x,\omega) = \oPhiin^+(\omega) \ee^{-\ii(\omega/v)x} + \oPhiout^+(\omega) \ee^{\ii(\omega/v)x}.
\end{equation}
By using the this expression,
Eq.~\eqref{eq:motion_sigma_w_app} is rewritten as
\begin{align} \label{eq:motion_q_sigma} % !!!!!!!!!!!!!!!!!!!!!!!!!!!!!!!!!!!!!!!!!!!!!!!
\frac{\dd}{\dd t}\osigma_{\mu,\nu}
& \simeq - \ii\omega_{\nu,\mu}\osigma_{\mu,\nu}
  + \left[ \osigma_{\mu,\nu}, \ophi \right]
    \int_0^{\infty}\dd\omega\ \ee^{-\ii\omega t}
      \frac{\ii\omega\omega_{\nu,\mu}\CC}{\hbar}
\nonumber \\ & \quad \times
      \left[\ophi^+(\omega)-\oPhiin^+(\omega) - \oPhiout^+(\omega) \right].
\end{align}
Further, the boundary condition \eqref{eq:boundary_LC-TL} is expressed as
\begin{equation} \label{eq:inout_Langevin_LC-TL} % !!!!!!!!!!!!!!!!!!!!!!!!!!!!!!!!!!!!!!!!!!!!
\ophi^+(\omega) - \oPhiin^+(\omega) - \oPhiout^+(\omega)
= - \ii\varLambda(\omega)\left[ \oPhiin^+(\omega) - \oPhiout^+(\omega) \right],
\end{equation}
where the non-dimensional value $\varLambda(\omega)$
characterizes the degree of confinement of the resonator:
\begin{equation} \label{eq:Lambda} % !!!!!!!!!!!!!!!!!!!!!!!!!!!!!!!!!!!!!!!!!
\varLambda(\omega) = \frac{1}{\omega\ZT\CC}.
\end{equation}
Then, the outgoing field is represented as
\begin{equation} \label{eq:oPhiout=phi-oPhiin} % !!!!!!!!!!!!!!!!!!!!!!!!!!!!!!!!
\oPhiout^+(\omega) = \frac{\ophi^+(\omega) - [1-\ii\varLambda(\omega)]\oPhiin^+(\omega)}{1 + \ii\varLambda(\omega)}.
\end{equation}
Substituting Eq.~\eqref{eq:oPhiout=phi-oPhiin} into Eq.~\eqref{eq:motion_q_sigma},
since we can focus on the narrow frequency region around the resonance
$\omega \sim \omega_{\nu,\mu}$ in the good-cavity limit, we get
\begin{align}
\frac{\dd}{\dd t}\osigma_{\mu,\nu}
& \simeq - \ii\omega_{\nu,\mu}\osigma_{\mu,\nu}
  + \left[ \osigma_{\mu,\nu}, \ophi \right]
    \int_0^{\infty}\dd\omega\ \ee^{-\ii\omega t}
      \frac{\ii\omega^2\CC}{\hbar}
\nonumber \\ & \quad \times
    \left[
        \frac{\ii\varLambda(\omega)}{1+\ii\varLambda(\omega)}\ophi^+(\omega)
      - \frac{\ii2\varLambda(\omega)}{1+\ii\varLambda(\omega)} \oPhiin^+(\omega)
    \right].
\end{align}
Further, we also get $\varLambda \gg 1$ ($\CC \ll 1/\omega\ZT$)
in the good-cavity limit,
and then this equation is approximated as
\begin{align} \label{eq:motion-sigma_Phiin} % !!!!!!!!!!!!!!!!!!!!!!!!!!!!!
\frac{\dd}{\dd t}\osigma_{\mu,\nu}
& \simeq - \ii\omega_{\nu,\mu}\osigma_{\mu,\nu}
  - \left[ \osigma_{\mu,\nu}, \ophi \right]
    \int_0^{\infty}\dd\omega\ \ee^{-\ii\omega t}
\nonumber \\ & \quad \times
      \left[
        \frac{\kappaLCz}{2} \left(\frac{\omega}{\wz}\right)^3
        \frac{2}{\hbar\ZR} \ophi^+(\omega)
      + \frac{\ii2\omega^2\CC}{\hbar} \oPhiin^+(\omega)
    \right].
\end{align}
Here, when the environment is large enough compared
with the resonator systems,
the incoming and outgoing fields corresponds to
the input and output operators, respectively, appearing in the Langevin equation
and the input-output relation \cite{Bamba2013MBC}
(corresponding to the Born-Markov approximation
performed in Sec.~\ref{sec:review}):
\begin{subequations}
\begin{align}
\oPhiin^+(\omega) & = \sqrt{\frac{\hbar\ZT}{2\omega}} \oain(\omega), \\
\oPhiout^+(\omega) & = \sqrt{\frac{\hbar\ZT}{2\omega}} \oaout(\omega).
\end{align}
\end{subequations}
Then, the equation of motion is finally rewritten as
\begin{subequations}
\begin{align}
\frac{\dd}{\dd t}\osigma_{\mu,\nu}
& \simeq - \ii\omega_{\nu,\mu}\osigma_{\mu,\nu}
  - \left[ \osigma_{\mu,\nu}, \ophi \right]
    \int_0^{\infty}\dd\omega\ \ee^{-\ii\omega t}
\nonumber \\ & \quad \times
      \left[
        \frac{\kappaLCz}{\hbar\ZR} \left(\frac{\omega}{\wz}\right)^3 \ophi^+(\omega)
      + \ii\sqrt{\frac{2\kappaLCz}{\hbar\ZR}\left(\frac{\omega}{\wz}\right)^3}\oain(\omega)
    \right].
\end{align}
On the other hand, in the good cavity limit,
Eq.~\eqref{eq:oPhiout=phi-oPhiin} is approximated as
\begin{equation}
\oaout(\omega) \simeq \oain(\omega)
- \ii\sqrt{\frac{2\kappaLCz}{\hbar\ZR}\left(\frac{\omega}{\wz}\right)^3}
  \ophi^+(\omega).
\end{equation}
\end{subequations}
These two equations correspond to the quantum Langevin equation \eqref{eq:Langevin_BornMarkov}
and input-output relation \eqref{eq:input-output_X}, respectively,
and this means that the SEC Hamiltonian is represented
in the good-cavity and independent-transition limit as
\begin{subequations} \label{eq:oHSE-general} % !!!!!!!!!!!!!!!!!!!!!!!!!!!!!!!!!!!
\begin{align}
\oHSE
& = \int_0^{\infty}\dd\omega\
  \hbar\sqrt{\frac{\kappaLCz}{2\pi}\left(\frac{\omega}{\wz}\right)^3\frac{2}{\hbar\ZR}}
  \oalphad(\omega) \ophi^+ + \Hc \\
& = \int_0^{\infty}\dd\omega\
  \hbar\sqrt{\frac{\kappaLCz}{2\pi}\left(\frac{\omega}{\wz}\right)^3}
  \oalphad(\omega) \left(\oa+\oad\right)^+ + \Hc
\end{align}
\end{subequations}
This expression is applicable to the LC-resonator
embedding any circuit elements even for $\partial\LA/\partial\dot{\phi} \neq 0$,
i.e., to both of the two circuits in Fig.~\ref{fig:1}.
However, we must pay attention that 
the Hamiltonians $\oHz$ of resonator systems must be derived under the same definition
of $\phi$ as drawn in Fig.~\ref{fig:1},
i.e., Eqs.~\eqref{eq:Hz_A} and \eqref{eq:Hz_B} should be used
for circuits A and B, respectively,
but the combination of Eq.~\eqref{eq:oHSE-general}
and $U^*\Hz^BU$ \eqref{eq:HzA=UHzBU} is not appropriate for circuit B.
In other words, under the same definition of $\phi$,
we can extend $\oHz$ to sophisticated systems
including anharmonicities and more than one CPBs.
Whereas the expression \eqref{eq:oHSE-general}
is independent of the detail of the resonator systems,
the loss rate from eigen-state $\ket{\nu}$ to $\ket{\mu}$
includes the information of the detail of resonator systems
through the matrix element $\braket{\mu|\ophi|\nu}$:
\begin{subequations} \label{eq:kappaLC_motion} % !!!!!!!!!!!!!!!!!!!!!!!!!!!!!!!!!
\begin{align}
\kappaLC_{\nu,\mu}
& = \kappaLCz \left(\frac{\omega_{\nu,\mu}}{\wz}\right)^3
    \frac{2}{\hbar\ZR} |\braket{\mu|\ophi|\nu}|^2 \\
& = \kappaLCz \left(\frac{\omega_{\nu,\mu}}{\wz}\right)^3
  |\braket{\mu|\oa+\oad|\nu}|^2.
\end{align}
\end{subequations}

\subsection{Comparison} \label{sec:compare_general}
The SEC Hamiltonian $\oHSE$ \eqref{eq:oHSE-general}
derived above is expressed by the flux $\ophi$ (not by $\oq$) in the resonator,
but the SEC is originally mediated by its time-derivative $\dot{\phi}$
(capacitive) as the second term in the Lagrangian \eqref{eq:L_A-LC-TL}.
In contrast, in the straightforward derivation, the SEC Hamiltonians $\oHSE^{A,B}$ are
derived as Eq.~\eqref{eq:oHSE-LCB-flux} and \eqref{eq:oHSE-A},
and they are rewritten in the good-cavity limit
(applying the pre-trace RWA) as
\begin{subequations} \label{eq:oHSE_AB_good} % !!!!!!!!!!!!!!!!!!!!!!!!!!!!!!!
\begin{align}
\oHSE^A
& \simeq \int_0^{\infty}\dd\omega\
    \ii\hbar\sqrt{\frac{\kappaLCz}{2\pi}\frac{\omega}{\wz}\frac{2\ZR}{\hbar}}
    \oalphad(\omega)\oq^+  + \Hc, \label{eq:oHSE_A_good} \\
\oHSE^B
& \simeq \int_0^{\infty}\dd\omega\
    \ii\hbar\sqrt{\frac{\kappaLCz}{2\pi}\frac{\omega}{\wz}\frac{2\ZR}{\hbar}}
\nonumber \\ & \quad \times
    \oalphad(\omega)
    \left[ \oq^+ + \frac{\CR}{\CJ}(\oq^++\orho^+) \right]
  + \Hc \label{eq:oHSE_B_good}
\end{align}
\end{subequations}
If the resonator is empty or the LC-resonator couples
with other elements inductively such as circuit A,
$\dot{\phi}$ corresponds to the charge in the resonator:
\begin{subequations} \label{eq:dot_phi=} % !!!!!!!!!!!!!!!!!!!!!!!!!!!!!!!!!!!
\begin{equation}
\CR \dot{\phi} = q.
\quad \text{(for $\partial \LA/\partial \dot{\phi} = 0$)}
\end{equation}
However, this relation is not generally valid
if the LC-resonator couples with other elements
capacitively ($\partial \LA/\partial\dot{\phi} \neq 0$),
and $\dot{\phi}$ is derived for circuit B as
\begin{equation}
\CR \dot{\phi}
= q + \frac{\CR}{\CJ} \left( q + \rho \right). \quad \text{(circuit B)}
\end{equation}
\end{subequations}
Here, comparing the above equations and the general expression \eqref{eq:oHSE-general},
we can find that Eqs.~\eqref{eq:oHSE_AB_good} are obtained from Eq.~\eqref{eq:oHSE-general}
by replacing
\begin{subequations} \label{eq:replace} % !!!!!!!!!!!!!!!!!!!!!!!!!!!!!!!!!!!!
\begin{align}
-\ii\omega\CR \ophi & \rightarrow \oq, \quad \text{(for $\partial \LA/\partial \dot{\phi} = 0$)} \\
-\ii\omega\CR \ophi & \rightarrow \oq + \frac{\CR}{\CJ} \left( \oq + \orho \right). \quad \text{(circuit B)}
\end{align}
\end{subequations}
This replacement is justified from Eqs.~\eqref{eq:dot_phi=}
and because we can focus the narrow frequency regions around the resonances
($\omega \sim \omega_{\nu,\mu}$) in the good-cavity and independent-transition limit.
Whereas it is not easy to check the validity of this replacement rigorously,
we can find this relation between the general (but approximated) expression \eqref{eq:oHSE-general}
and straightforward ones \eqref{eq:oHSE_AB_good}
at least for the SEC considered in this paper.
When the transitions of interest are not well isolated with each other
or the quality of the resonator is not good enough,
we cannot use the general one \eqref{eq:oHSE-general},
but we should rather consider Eqs.~\eqref{eq:oHSE_AB_good},
which can be derived by the above replacement
even for other structures of circuits.
Then, the SEC Hamiltonians need not be derived for each circuits of interest
by the straightforward way,
but the simple replacement \eqref{eq:replace} is efficient for deriving them.

In the case of $\partial \LA/\partial \dot{\phi} = 0$,
the SEC Hamiltonian is expressed as Eq.~\eqref{eq:oHSE-general}
and also as Eq.~\eqref{eq:oHSE_A_good},
which corresponds to $\oHSE^+$ and $\oHSE^-$ in Eq.~\eqref{eq:oHSE_pm}, respectively.
Then, the loss rate from state $\ket{\nu}$ to $\ket{\mu}$
can be expressed in the following two ways
($\partial\LA/\partial\dot{\phi} = 0$)
\begin{align}
\kappa_{\nu,\mu}^{A}
& = \kappaLCz \left(\frac{\omega_{\nu,\mu}}{\wz}\right)^3
    \left|\braket{\mu|\oa+\oad|\nu}\right|^2
\nonumber \\ 
& = \kappaLCz \left(\frac{\omega_{\nu,\mu}}{\wz}\right)
    \left|\braket{\mu|\oa-\oad|\nu}\right|^2.
\end{align}
Since the eigen-frequencies are strongly shifted from the bare ones $\wz$ and $\wx$
in the ultrastrong light-matter interaction regime,
the factors $(\omega/\wz)$ and $(\omega/\wz)^3$ are quite important
for considering the dissipation correctly.
However, ignoring the $(\omega /\omega _0)$-factor,
the simple SEC Hamiltonians \eqref{eq:oHSE_pm}
can be justified when we suppose the circuits
with $\partial \LA/\partial \dot{\phi} = 0$.
In experiments of superconducting circuits,
the ``inductive'' ultrastrong interaction has been realized
\cite{Niemczyk2010NP,Fedorov2010PRL,Forn-Diaz2010PRL},
and it corresponds to the situation of $\partial \LA/\partial \dot{\phi} = 0$.
Then, we can use Eq.~\eqref{eq:oHSE-general} or \eqref{eq:oHSE_A_good}
for the SEC Hamiltonian.
However, for ``capacitive'' ultrastrong interaction
with the capacitive SEC (as in Fig.~\ref{fig:1})
or for the combination of inductive ultrastrong interaction and inductive SEC
(we cannot find any restrictions to prevent them at least
in the discussion of this paper)
we should use the general one \eqref{eq:oHSE-general}
or complicated expressions such as Eq.~\eqref{eq:oHSE_B_good}
must be derived for given circuits.
In any cases, the SEC Hamiltonian $\oHSE$
and $\oHz$ of resonator systems must be derived
under the same definition of resonator field $\phi$.
Otherwise, we get incorrect loss rates.

\subsection{Demonstration} \label{sec:demo_general}
In the good-cavity and independent-transition limit,
the SEC Hamiltonian is derived generally as Eq.~\eqref{eq:oHSE-general},
which is applicable to both circuits A and B under the definition of $\phi$
as depicted in Fig.~\ref{fig:1}.
Then, the Hamiltonians $\oHz^{A,B}$ are expressed as
Eqs.~\eqref{eq:oHz-L-C} when we neglect the anharmonicity of the CPB.
In the same manner as Sec.~\ref{sec:demo_straight},
the SEC Hamiltonian \eqref{eq:oHSE-general} is rewritten as
\begin{align}
\oHSE & = \int_0^{\infty}\dd\omega\
  \hbar\sqrt{\frac{\kappaLCz}{2\pi}\left(\frac{\omega}{\wz}\right)^3} \oalphad(\omega)
\nonumber \\ & \quad \times
  \sum_{j=L,U}\left(w_j^{\xi}-y_j^{\xi}\right)^*\op_j^{\xi} + \Hc
\end{align}
Then, the loss rates of eigen-modes $j=L,U$ are expressed for each circuit $\xi=A,B$ as
\begin{equation}
\kappa_j^{\xi}
= \kappaLCz \left(\frac{\omega_j}{\wz}\right)^3
  \left|w_j^{\xi}-y_j^{\xi}\right|^2.
\end{equation}
We can check that this expression certainly reproduces all the results
in Fig.~\ref{fig:2}, which are obtained from the SEC Hamiltonians $\oHSE^{A,B}$
\eqref{eq:oHSE-LCB-flux} and \eqref{eq:oHSE-A}
derived independently for circuits A and B, respectively,
by the straightforward calculation.
Although the SEC Hamiltonians are derived in different forms
depending on the detail of resonator systems
by the straightforward calculation,
Eq.~\eqref{eq:oHSE-general} remains its expression for any circuit elements
in the LC-resonator, if the SEC itself is not changed.
When we want to describe circuit B by the same system-of-interest Hamiltonian
as $\oHz^A = \oUd\oHz^B\oU$, the SEC Hamiltonian \eqref{eq:oHSE-general}
is rewritten as
\begin{align} \label{eq:UdHSEU} % !!!!!!!!!!!!!!!!!!!!!!!!!!!!!!!!!!!!!!!!!!!!
\oUd\oHSE\oU & = \int_0^{\infty}\dd\omega\
  \hbar\sqrt{\frac{\kappaLCz}{2\pi}\left(\frac{\omega}{\wz}\right)^3\frac{2}{\hbar\ZR}}
\nonumber \\ & \quad \times
  \left[ \oalphad(\omega) \left(\ophi^+ + \opsi^+\right) + \Hc \right].
\end{align}
Then, the loss rates in circuit B are also expressed as
\begin{equation}
\kappa^B_j = \kappaLCz\left(\frac{\omega_j}{\wz}\right)^3
\left| (w^A_j-y^A_j) + 2\grabi(x^A_j-z^A_j) \right|^2.
\end{equation}
This also gives the same loss rates depicted in Fig.~\ref{fig:2}.
In this way, we eventually get different Hamiltonians as a whole
for circuits A and B, and the dissipative behaviors are different
especially in the ultrastrong light-matter interaction regime
as demonstrated in Figs.~\ref{fig:2}(d) and (f).
In addition, since the SEC Hamiltonian can be transformed by the unitary operator $\oU$
as above,
we must pay an attention the definition of the flux $\phi$ of the LC-resonator
as Fig.~\ref{fig:1},
under which the SEC Hamiltonian \eqref{eq:oHSE-general} is derived.
In other words, $\oHz$ and $\oHSE$ must be derived correctly
under the same definition of $\phi$.

\section{Fabry-Perot cavity} \label{sec:FP}
\begin{figure}[tbp]
\begin{center}
\includegraphics[width=.8\linewidth]{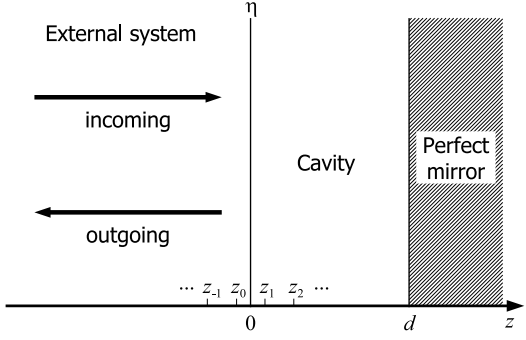}
\end{center}
\caption{Sketch of a Fabry-Perot cavity. A perfect mirror is placed at $z = \cavlen$
and a non-perfect mirror is at $z = 0$.}
\label{fig:3}
\end{figure}
Whereas we discuss the SEC in the system of superconducting circuits
in previous sections,
we can discuss the SEC also for other systems
starting from Lagrangians describing them.
We next derive the SEC Hamiltonian for a Fabry-Perot cavity
confining the electromagnetic fields.
As supposed in our previous work \cite{Bamba2013MBC},
we also consider the cavity consisting of a perfect mirror
and a non-perfect one as depicted in Fig.~\ref{fig:3}.

We first discuss the Lagrangian describing the electromagnetic fields
interacting with matters in Sec.~\ref{sec:Lagrangian-EM}.
Then, in Sec.~\ref{sec:SEC-FP},
we derive the SEC Hamiltonian,
by which the results obtained in Ref.~\cite{Bamba2013MBC}
can be certainly reproduced.
We also discuss some prospects for the SECs of split-ring resonators
and of subwavelength structures in Sec.~\ref{sec:SRR}.

\subsection{Lagrangian for electromagnetic fields} \label{sec:Lagrangian-EM}
We first consider general systems of the electromagnetic fields
interacting with matters.
Here, we take the Coulomb gauge and introduce the vector potential $\vA(\vr)$
giving the transverse electric field $\vET(\vr)$ and the magnetic flux $\vB(\vr)$
as
\begin{subequations}
\begin{align}
\vET(\vr) & = - \dot{\vA}(z), \\
\vB(\vr) & = \rot\vA(\vr).
\end{align}
\end{subequations}
From these relations, we directly get the following Maxwell's equation
\begin{equation} \label{eq:Maxwell_rotE=dBdt} % !!!!!!!!!!!!!!!!!!!!!!!!!!!!!!
\rot\vE(\vr) = - \dot{\vB}(\vr),
\end{equation}
and the magnetic flux automatically satisfy
\begin{equation} \label{eq:divB=0} % !!!!!!!!!!!!!!!!!!!!!!!!!!!!!!!!!!!!!!!!
\div\vB(\vr) = 0.
\end{equation}
The longitudinal component of the electric field $\vEL(\vr) = - \grad\scp(\vr)$
is expressed by the scalar potential $\scp(\vr)$.
As discussed in Refs.~\cite{cohen-tannoudji89,Glauber1991PRA,knoll01},
the Lagrangian of the total system is represented as $\LEB + \LEBM$,
where the matter and the light-matter interaction is described by $\LEBM$,
and $\LEB$ is the Lagrangian representing
the transverse electric field $\vET(\vr)$ and the magnetic flux $\vB(\vr)$ as
\begin{equation}
\LEB = \frac{1}{2} \int\dd\vr\left[ \diez\diebg(\vr)\vET(\vr)^2 - \frac{\vB(\vr)^2}{\muz} \right].
\end{equation}
Here, $\diebg(\vr)$ is the background dielectric constant
except the matters of interest, and it depends on position $\vr$
\cite{Glauber1991PRA,knoll01}.
$\diez$ and $\muz$ are the vacuum permittivity and permeability, respectively.
In terms of $\vA(\vr)$, the Lagrangian is rewritten as
\begin{equation}
\LEB = \frac{1}{2} \int\dd\vr\left\{
    \diez\diebg(\vr)\dot{\vA}(\vr)^2
  - \frac{\left[\rot\vA(\vr)\right]^2}{\muz}
  \right\}.
\end{equation}
Then, the Lagrange equation with respect to $\vA(\vr)$ is derived as
\begin{align}&
- \diez\diebg(\omega)\dot{\vE}{\perp}(\vr) + \ddt{}\frac{\partial \LEBM}{\partial \dot{\vA}(z)}
\nonumber \\ & \quad
= - \frac{\rot\vB(\vr)}{\muz} + \frac{\pbar \LEBM}{\pbar \vA(\vr)},
\end{align}
where we define
\begin{equation}
\frac{\pbar \LEBM}{\pbar A_{\xi}(\vr)}
= \frac{\partial \LEBM}{\partial A_{\xi}(\vr)}
- \div \frac{\partial \LEBM}{\partial(\grad A_{\xi}(\vr))}.
\end{equation}
Here, we introduce the electric displacement field $\vD(\vr)$ and the magnetic field $\vH(\vr)$ as
\begin{subequations}
\begin{align}
\vD(\vr) & = \diez\diebg(\vr)\vE(\vr) + \vP(\vr), \\
\vH(\vr) & = \vB(\vr) / \muz - \vM(\vr),
\end{align}
\end{subequations}
where $\vP(\vr)$ is the polarization density and $\vM(\vr)$ is the magnetization.
We suppose globally neutral systems and there is no external current,
and the displacement field satisfies
\begin{equation} \label{eq:divD=0} % !!!!!!!!!!!!!!!!!!!!!!!!!!!!!!!!!!!!!!!!!
\div\vD(\vr) = 0.
\end{equation}
Then, the above Lagrange equation is rewritten as
\begin{align}&
-\dot{\vD}(\vr) + \dot{\vP}_{\perp}(\vr) + \ddt{}\frac{\partial \LEBM}{\partial \dot{\vA}(\vr)}
\nonumber \\ & \quad
= - \rot[\vH(\vr)+\vM(\vr)] + \frac{\pbar \LEBM}{\pbar \vA(\vr)}.
\end{align}
In the absence of external current,
the two fields $\vD(\vr)$ and $\vH(\vr)$ must satisfy
\begin{equation} \label{eq:Maxwell_rotH=dDdt} % !!!!!!!!!!!!!!!!!!!!!!!!!!!!!!
\dot{\vD}(\vr) = \rot\vH(\vr).
\end{equation}
Then, from the above Lagrange equation,
the polarization and magnetization should have the following relation
\begin{equation}
\dot{\vP}_{\perp}(\vr) + \ddt{}\frac{\partial \LEBM}{\partial \dot{\vA}(\vr)}
= - \rot\vM(\vr) + \frac{\pbar \LEBM}{\pbar \vA(\vr)}.
\end{equation}

On the other hand, due to the duality of the electric and magnetic fields,
we can also describe the Lagrangian in terms of $\vD(\vr)$
and $\vH(\vr)$ for globally neutral systems.
The total Lagrangian is represented as $\LDH + \LDHM$,
and $\LDH$ is expressed as
\begin{equation} \label{eq:LR-D-P-H-M} % !!!!!!!!!!!!!!!!!!!!!!!!!!!!!!!!!!!!!
\LDH
= \frac{1}{2} \int\dd\vr\left\{
    \muz \vHT(\vr)^2 - \frac{\vD(\vr)^2}{\diez\diebg(\vr)}
  \right\}.
\end{equation}
Here, we introduce a new transverse field $\vY(z)$ giving
\begin{subequations}
\begin{align}
\vHT(\vr) & = \dot{\vY}(\vr), \\
\vD(\vr) & = \rot\vY(\vr).
\end{align}
\end{subequations}
By this definition, the two Maxwell's equations
\eqref{eq:divD=0} and \eqref{eq:Maxwell_rotH=dDdt}
are automatically satisfied.
Further, since Eq.~\eqref{eq:divB=0} should also be satisfied,
the longitudinal components of the magnetic field and the magnetization
are related as $\vHL(\vr) = - \vM_{\parallel}(\vr)$.
In terms of $\vY(\vr)$, the Lagrangian $\LDH$ is rewritten as
\begin{equation}
\LDH
= \frac{1}{2} \int\dd\vr\left\{
    \muz \dot{\vY}(\vr)^2 
  - \frac{[\rot\vY(\vr)]^2}{\diez\diebg(\vr)}
  \right\}.
\end{equation}
Then, from the total Lagrangian $\LDH+\LDHM$,
the Lagrange equation with respect to $\vY(\vr)$ is derived as
\begin{equation}
\dot{\vH}_{\perp}(\vr) + \ddt{}\frac{\partial \LDHM}{\partial \dot{\vY}(\vr)}
= - \rot\frac{\vD(\vr)}{\diez\diebg(\vr)}
  + \frac{\pbar \LDHM}{\pbar \vY(\vr)}.
\end{equation}
Since the electric field and the magnetic flux satisfy
Eq.~\eqref{eq:Maxwell_rotE=dBdt},
the magnetization and the polarization must have the following relation in this case
\begin{equation}
- \dot{\vM}_{\perp}(\vr) + \ddt{}\frac{\partial \LDHM}{\partial \dot{\vY}(\vr)}
= - \rot\frac{\vP(\vr)}{\diez\diebg(\vr)}
    + \frac{\pbar \LDHM}{\pbar \vY(\vr)}.
\end{equation}
Of course, the two Lagrangians
$\LEB + \LEBM$ and $\LDH + \LDHM$
are equivalent in the sense that both of them certainly give
the Maxwell's equations and equations of motion of charged particles
in matters \cite{cohen-tannoudji89}.

\subsection{SEC for Fabry-Perot cavity} \label{sec:SEC-FP}
We next derive the SEC Hamiltonian for the Fabry-Perot cavity
depicted in Fig.~\ref{fig:3} \cite{Lang1973PRA,Bamba2013MBC}.
The system is terminated by the perfect mirror at $z = \cavlen$,
and a non-perfect thin mirror is placed at $z = 0$.
The mirror is described by the background dielectric constant
as $\diebg(z) = \eta\delta(z)$,
and $\eta$ determines the degree of confinement of the cavity fields
(and the reflectivity of the non-perfect mirror).
The electromagnetic fields are confined in $0 < z < \cavlen$,
but the confinement is not perfect through the non-perfect mirror at $z = 0$.
Atoms or matters exist inside the cavity structure.
For a good correspondence with the discussion in Sec.~\ref{sec:derive_SEC},
we take the Lagrangian \eqref{eq:LR-D-P-H-M}
described by the displacement field $D(z)$ and the magnetic field $H(z)$.
In the discrete description as depicted in Fig.~\ref{fig:3},
the Lagrangian is expressed as
\begin{align} \label{eq:LR-Fabry-Perot} % !!!!!!!!!!!!!!!!!!!!!!!!!!!!!!!!!
\LDH
& = \frac{1}{2} \left\{
    \sum_{j} \muz\dot{\YY}(z_j)^2\dz
  - \frac{[\YY(z_1)-\YY(z_{0})]^2}{\diez\eta}
\right. \nonumber \\ & \quad \left.
  - \sum_{j\neq1}
    \frac{[\YY(z_j)-\YY(z_{j-1})]^2}{\diez\dz}
  \right\}.
\end{align}
Here, $\dz$ is the distance between the discrete positions,
$z_1 = \dz / 2$, and $z_0 = - \dz / 2$.
The cavity system is described by $j \geq 1$,
external system is by $j \leq 0$,
and the second term in Eq.~\eqref{eq:LR-Fabry-Perot} corresponds to the coupling between them.
From this Lagrangian, we get the following relation
in the discrete and continuous base:
\begin{subequations}
\begin{align}
\frac{\partial \LDH}{\partial \YY(z_1)}
& = - \frac{\YY(z_1)-\YY(z_2)}{\diez\dz}
  - \frac{\YY(z_1)-\YY(z_0)}{\diez\eta}, \\
\frac{\partial \LDH}{\partial \YY(0^+)}
& = \frac{1}{\diez}\left.\ddz{\YY}\right|_{z=0^+}
  - \frac{\YY(0^+)-\YY(0^-)}{\diez\eta}.
\label{eq:dLRdX} % !!!!!!!!!!!!!!!!!!!!!!!!!!!!!!!!!!!!!!!!!!!!!!!!!!!!!!!!!!!
\end{align}
\end{subequations}
The first term gives the boundary condition in the case of a perfect mirror,
i.e., $D(z=0^+) = 0$.
The other terms represent the connection with the external field,
and it is decreased with the increase of $\eta$.
Here, we suppose that the cavity system including the matters inside
is properly diagonalized in the case of the perfect cavity.
We denote the eigen-states as $\{\ket{\mu}\}$ and the eigen-frequencies as $\{\omega_{\mu}\}$.
From the total Lagrangian $L = \LDH + \LDHM$ or its Hamiltonian,
since the cavity field connects with the external system only through
the second term of Eq.~\eqref{eq:LR-Fabry-Perot} (or the one of Eq.~\eqref{eq:dLRdX}),
the equation of motion of $\osigma_{\mu,\nu} = \ket{\mu}\bra{\nu}$ is derived
in the good-cavity and independent-transition limit as
\begin{subequations}
\begin{align}
\ddt{}\osigma_{\mu,\nu}
& = - \ii\omega_{\nu,\mu}\osigma_{\mu,\nu}
    + \frac{1}{\ii\hbar}\frac{1}{2\diez\eta}\left[ \osigma_{\mu,\nu}, \{\oY(z_1)-\oY(z_0)\}^2 \right] \\
& = - \ii\omega_{\nu,\mu}\osigma_{\mu,\nu}
    + \left[ \osigma_{\mu,\nu}, \oY(0^+) \right]
      \frac{\oY(0^+)-\oY(0^-)}{\ii\hbar\diez\eta} \\
& \simeq - \ii\omega_{\nu,\mu}\osigma_{\mu,\nu}
    + \left[ \osigma_{\mu,\nu}, \oH(0^+) \right]
\nonumber \\ & \quad \times
      \int_0^{\infty}\dd\omega\
      \frac{\ee^{-\ii\omega t}}{\ii\hbar\diez\eta\omega^2}
      \left[ \oH^+(0^+,\omega) - \oH^+(0^-,\omega) \right].
\label{eq:motion-sigma-MBC} % !!!!!!!!!!!!!!!!!!!!!!!!!!!!!!!!!!!!!!!!!!!!!!!!
\end{align}
\end{subequations}
Here, from the integral from of Eq.~\eqref{eq:Maxwell_rotH=dDdt},
we get the following Maxwell's boundary condition \cite{Bamba2013MBC}
\begin{equation} \label{eq:MBC-HE} % !!!!!!!!!!!!!!!!!!!!!!!!!!!!!!!!!!!!!!!!!
H(0^-,\omega) - H(0^+,\omega)
= - \ii\omega\diez\eta E(0^-,\omega).
\end{equation}
Since the electromagnetic fields obey the simple wave equation
in the external system, the external field in $z < 0$ can be rewritten
by the incoming field $\oHin$ and the outgoing one $\oHout$ as
\begin{equation} \label{eq:H-in-out} % !!!!!!!!!!!!!!!!!!!!!!!!!!!!!!!!!!!!!!!
\oH(z<0,\omega) = \ee^{\ii(\omega/c)z} \oHin(\omega) + \ee^{-\ii(\omega/c)z} \oHout(\omega).
\end{equation}
From the boundary condition \eqref{eq:MBC-HE},
the outgoing field is expressed as
\begin{equation}
\oHout(\omega) = \frac{\oH(0^+,\omega) - [1+\ii\varLambda(\omega)]\oHin(\omega)}{1-\ii\varLambda(\omega)},
\end{equation}
where $\varLambda(\omega) = \omega\eta/c$.
Substituting this into Eq.~\eqref{eq:motion-sigma-MBC},
in the good cavity limit ($\varLambda \gg 1$), we get
\begin{align} \label{eq:motion-sigma-Hin} % !!!!!!!!!!!!!!!!!!!!!!!!!!!!!!!!
\ddt{}\osigma_{\mu,\nu}
& \simeq - \ii\omega_{\nu,\mu}\osigma_{\mu,\nu}
  - \left[ \osigma_{\mu,\nu}, \oH(0^+) \right]
    \int_0^{\infty}\dd\omega\ \ee^{-\ii\omega t}
\nonumber \\ & \quad \times
  \left[
      \frac{\muz c}{\hbar\omega\varLambda(\omega)^2} \oH^+(0^+,\omega)
    + \frac{2}{\ii\hbar\diez\omega c \varLambda(\omega)} \oHin \right].
\end{align}
Since the incoming and outgoing fields are represented as
\begin{subequations}
\begin{align}
\oHin(\omega) & = \frac{\ii\omega}{\muz c}\sqrt{\frac{\hbar}{2\diez c\omega}} \oain(\omega), \\
\oHout(\omega) & = \frac{\ii\omega}{\muz c}\sqrt{\frac{\hbar}{2\diez c\omega}} \oaout(\omega),
\end{align}
\end{subequations}
Eq.~\eqref{eq:motion-sigma-Hin} is rewritten as
\begin{subequations}
\begin{align}
\ddt{}\osigma_{\mu,\nu}
& \simeq - \ii\omega_{\nu,\mu}\osigma_{\mu,\nu}
  - \left[ \osigma_{\mu,\nu}, \oH(0^+) \right]
    \int_0^{\infty}\dd\omega\
      \ee^{-\ii\omega t}
\nonumber \\ & \quad \times
  \left[
      \frac{\muz c}{\hbar\omega\varLambda(\omega)^2} \oH^+(0^+,\omega)
    + \sqrt{\frac{2\muz c}{\hbar\omega\varLambda(\omega)^2}} \oain \right].
\end{align}
On the other hand, from Eq.~\eqref{eq:H-in-out}, we get
\begin{equation}
\oaout(\omega) \simeq \oain(\omega)
+ \sqrt{\frac{2\muz c}{\hbar\omega\varLambda(\omega)^2}} \oH^+(0^+,\omega).
\end{equation}
\end{subequations}
These two equations
correspond to the quantum Langevin equation and the input-output relation,
and this fact means that the SEC is expressed in this case as
\begin{equation} \label{eq:oHSE_FP} % !!!!!!!!!!!!!!!!!!!!!!!!!!!!!!!!!!!!!!!!
\oHSE = \int_0^{\infty}\dd\omega\ \ii\hbar\sqrt{\frac{\muz c}{\pi\hbar\omega\varLambda(\omega)^2}}
\left[ \oalphad(\omega) \oH^+(0^+) - \Hc \right].
\end{equation}

In the independent-transition limit,
the loss rate from $\ket{\nu}$ to $\ket{\mu}$ ($\mu < \nu$) is obtained as
\begin{equation}
\kappa_{\nu,\mu} = \frac{2\muz c}{\hbar\omega_{\nu,\mu}\varLambda(\omega_{\nu,\mu})^2}|\braket{\mu|\oH(0^+)|\nu}|^2.
\end{equation}
If there is no induced magnetization inside the cavity,
since $H = B/\muz$ and $\partial E/\partial z = - \dot{B}$, 
the loss rate is also rewritten as
\begin{subequations}
\begin{align}
\kappa_{\nu,\mu}
& = \frac{2c}{\hbar\muz\omega_{\nu,\mu}\varLambda(\omega_{\nu,\mu})^2}|\braket{\mu|\oB(0^+)|\nu}|^2 \\
& = \frac{2c}{\hbar\muz\omega_{\nu,\mu}{}^3\varLambda(\omega_{\nu,\mu})^2}
\left|\braket{\mu|\ddz{\oE}|\nu}\right|^2_{z=0^+}.
\end{align}
\end{subequations}
Here, for the cavity depicted in Fig.~\ref{fig:3},
the electric and magnetic fields are expressed as
\begin{subequations}
\begin{align}
\oE(z)
& = \sum_{m=1}^{\infty}\ii\omega_m
    \sqrt{\frac{\hbar}{\diez\omega_m\cavlen}}
    (\oa_m-\oad_m) \sin[(\omega_m/c)z], \\
\oB(z)
& = \sum_{m=1}^{\infty}\frac{\omega_m}{c}
    \sqrt{\frac{\hbar}{\diez\omega_m\cavlen}}
    (\oa_m+\oad_m) \cos[(\omega_m/c)z],
\end{align}
\end{subequations}
where $\omega_m = m(\pi c/\cavlen)$ for integer $m = 1, 2, \ldots$
If the eigen-states are mainly composed by the $m$-th cavity mode,
the loss rate is also represented as
\begin{subequations} \label{eq:kappa_FP} % !!!!!!!!!!!!!!!!!!!!!!!!!!!!!!!!!!!
\begin{align}
\kappa_{\nu,\mu}
& = \kappaFPz(\omega_{\nu,\mu})\left(\frac{\omega_m}{\omega_{\nu,\mu}}\right)
|\braket{\mu|\oa_m+\oad_m|\nu}|^2 \\
& = \kappaFPz(\omega_{\nu,\mu})\left(\frac{\omega_m}{\omega_{\nu,\mu}}\right)^3
|\braket{\mu|\oa_m-\oad_m|\nu}|^2,
\end{align}
\end{subequations}
where $\kappaFPz(\omega)$ is the loss rate of the bare cavity mode.
\begin{equation}
\kappaFPz(\omega) = \frac{2c}{\varLambda(\omega)^2\cavlen}.
\end{equation}
These expressions certainly reproduce the results in our previous work \cite{Bamba2013MBC},
where the Fabry-Perot cavity filled by the medium with bosonic excitations.

When the light-matter interaction is mediated by the electric polarization
and the electromagnetic fields are confined by the Fabry-Perot cavity
as Fig.~\ref{fig:3},
the loss rates are expressed as simply as Eq.~\eqref{eq:kappa_FP}.
This situation corresponds to circuit A in Fig.~\ref{fig:1},
where $\partial\LA/\partial\dot{\phi} = 0$.
Whereas the ultrastrong light-matter interaction has been realized
by the electric dipole transitions experimentally
\cite{Gunter2009N,Anappara2009PRB,Todorov2009PRL,Todorov2010PRL,Todorov2012PRB,Porer2012PRB,Schwartz2011PRL,Scalari2012S},
the situation corresponding t circuit B,
where the SEC Hamiltonians are derived in complicated forms
by the straightforward calculation,
can be realized
if the ultrastrong light-matter interaction is mediated
by the magnetic field.
Even for the electric dipole transitions,
the circuit-B situation could be realized for other cavity structures
such as the sprit-ring resonators \cite{Scalari2012S}
and the subwavelength structures
\cite{Todorov2009PRL,Todorov2010PRL,Todorov2012PRB,Porer2012PRB}.
The SEC Hamiltonians for these structures should have different forms.

\subsection{Split-ring resonators \& subwavelength structures} \label{sec:SRR}
In Ref.~\cite{Scalari2012S},
the transition between Landau levels in two-dimensional electron gas
is coupled ultrastrongly with the modes in the split-ring resonators.
The current with THz frequency in each resonator is similar to
the superconducting current in the LC-resonator in Fig.~\ref{fig:1}.
Then, the SEC of the split-rings could be discussed based on the Lagrangians
describing their equivalent circuit model \cite{Baena2005MTT}.
On the other hand, the split-rings are fabricated periodically on the sample \cite{Scalari2012S},
and they behave as a meta-material layer for the electromagnetic fields.
Then, we can also describe the sprit-ring resonators
as a resonator of the electromagnetic fields
in the meta-material with effective permittivity and permeability
(with strong frequency-dependence) as discussed in this section.
In this way, the split-ring resonators can be analyzed in such two manners.
However, a sophisticated analysis is in principle required
for describing the complicated structures of the split-rings.

The subwavelength structures \cite{Todorov2009PRL,Todorov2010PRL,Todorov2012PRB,Porer2012PRB}
enhance the confinement of the electromagnetic fields,
and then the light-matter interaction is also enhanced.
Further, owing to the photonic bands constructed by the periodic subwavelength structures,
we can access directly to the surface plasmons.
In order to analyze these kind of structures,
we must extend the simple Fabry-Perot cavity
to more complex three-dimensional systems.
Whereas the extension is in principle possible,
the approach of Ref.~\cite{Bamba2013MBC} is rather appropriate
for supposing such complicated structures
as the first step of the investigation,
because the excitations between the subbands in semiconductor quantum wells
are nearly bosonic
and the cavity structure is complicated.

\section{Summary} \label{sec:summary}
We derived the SEC Hamiltonians by the straightforward calculation
from the Lagrangians describing in detail the mechanisms of loss and confinement
of the cavity fields.
However, we found that the SEC Hamiltonians are in principle
modified by the presence of the interaction between the cavity fields (light)
and other ones (matters) inside the cavity.
When the quality of the cavity is high enough (good-cavity limit: $\kappa \ll \wz, \wx$),
the SEC Hamiltonians should be derived as follows
\begin{itemize}
\item
In the weak and normally strong light-matter interaction regimes ($\rabi \ll \wz, \wx$),
the SEC Hamiltonians can be reduced to the standard one, Eq.~\eqref{eq:oHSE_standard},
which can be derived also for empty cavities.
This is because the RWA can be applied to the light-matter interaction
in the basis of photons and excitations,
and the total number of photons and excitations is conserved.
\item
In the ultrastrong light-matter interaction regime ($\rabi \gtrsim \wz, \wx$)
and in the independent-transition limit,
the SEC Hamiltonians can be derived by the procedure in Sec.~\ref{sec:derive_SEC},
and the derived Hamiltonians such as Eqs.~\eqref{eq:oHSE-general} and \eqref{eq:oHSE_FP}
are generally valid independent of the detail of the light-matter interaction
inside the cavity.
However, we must pay attention that the Hamiltonians of cavity systems
must be derived under the same gauge for deriving the SEC ones
(under the same definition of resonator field $\phi$ as depicted in Fig.~\ref{fig:1}).
\item
In the ultrastrong light-matter interaction regime
but not in the independent-transition limit,
in which we cannot perform the post-trace RWA,
the SEC Hamiltonians should be derived by the straightforward calculation,
and the expressions such as \eqref{eq:oHSE-LCB-flux}, \eqref{eq:oHSE_A_a},
and \eqref{eq:oHSE-A} are obtained
depending on the detail of cavity systems in principle.
However, as far as we checked for simple systems depicted in Fig.~\ref{fig:1},
we can obtain the straightforward SEC Hamiltonians in the good-cavity limit
by the replacement \eqref{eq:replace}
from the general expression \eqref{eq:oHSE-general}.
\end{itemize}
In the bad-cavity limit, in which the pre-trace RWA cannot be applied,
we must also derive the SEC Hamiltonians by the straightforward way.
Further, for discussing not only the dissipation but also the Lamb shift
due to the coupling with the environment,
the straightforward SEC Hamiltonians are also desired.

In the ultrastrong light-matter interaction regime,
the SEC Hamiltonians have the additional degree of freedom
as seen in Eqs.~\eqref{eq:oHSE_pm}, i.e., whether the SEC is
mediated by the electric field (capacitive) or by the magnetic field (inductive).
It is determined by the mechanisms of the loss and confinement of the cavity field,
and the SEC Hamiltonians should be derived as explained above.
Even if the SEC and the cavity system are equivalent
as for circuits A and B in Fig.~\ref{fig:1},
the difference of the arrangement for the two circuits
gives the difference of the loss rates of the cavity systems
as demonstrated in Fig.~\ref{fig:2}.
In the independent-transition limit,
we can use the general expression $\oHSE$ \eqref{eq:oHSE-general}
of the SEC Hamiltonian for both circuits,
while the Hamiltonians of the cavity systems are derived
in different forms $\oHz^A$ \eqref{eq:Hz_A} and $\oHz^B$ \eqref{eq:Hz_B}.
When we use the same Hamiltonian
$\oHz^A = \oUd\oHz^B\oU$ for the cavity system,
the SEC Hamiltonians are transformed in different expressions
$\oHSE$ \eqref{eq:oHSE-general} and $\oUd\oHSE\oU$ \eqref{eq:UdHSEU}.
As the result by these differences,
the loss rates become significantly different for the two circuits
in the ultrastrong light-matter interaction regime
[Figs.~\ref{fig:2}(d) and (f)],
while the difference is negligible in the weak and normally strong ones
[Figs.~\ref{fig:2}(c) and (e)].

We also face the similar question whether the nonlinearities
and dissipation in matters are mediated by the electric polarization, magnetization,
or just the number of excitations.
Whereas the ultrastrong interactions and good cavities can be realized
only in limited structures currently,
there is a possibility for controlling the inevitable dissipation
by the ultrastrong interactions
with totally modeling the light-matter interaction and the SEC.

\begin{acknowledgments}
M.~B.~thanks to Pierre Nataf and Makoto Yamaguchi for fruitful discussions.
This work was supported by KAKENHI 24-632
and the JSPS through its FIRST Program.
\end{acknowledgments}

\appendix
\section{SEC for transmission line resonator} \label{sec:TLR-TL}
\begin{figure*}[tbp]
\includegraphics[width=.8\linewidth]{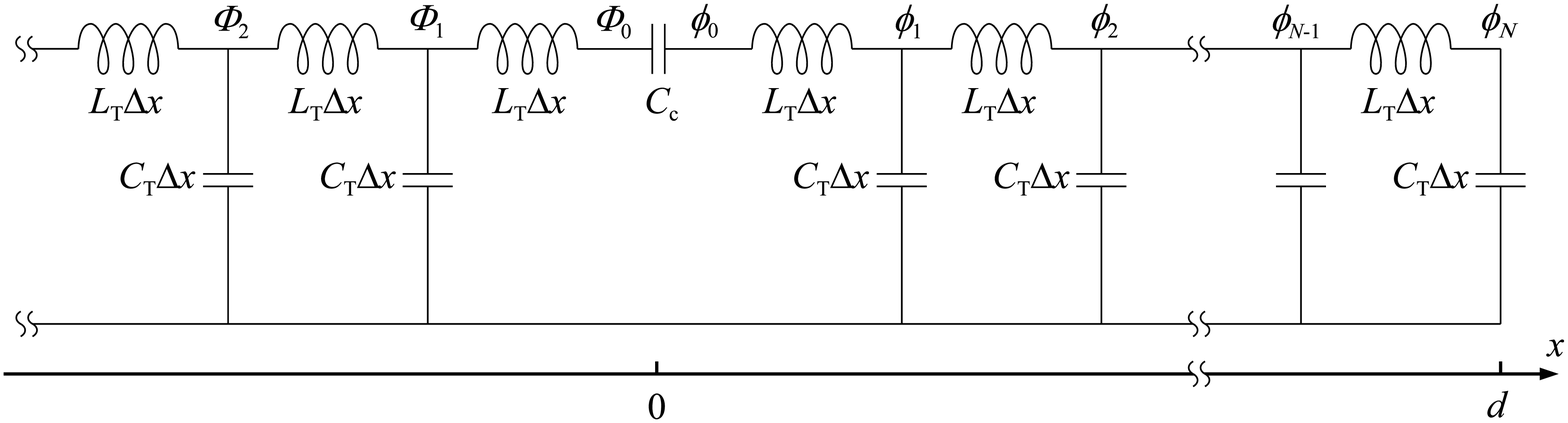}
\caption{A resonator is made by a transmission line with a finite length $\cavlen$.
It couples with a semi-infinite transmission line by capacitance $\CC$.
Artificial atoms can exist inside the resonator
except at the contact with the external line.}
\label{fig:TLR-TL}
\end{figure*}
Instead of the LC-resonator, here we consider a resonator
fabricated by a transmission line with a finite length $\cavlen$
as depicted in Fig.~\ref{fig:TLR-TL}.
The total Lagrangian is represented as
\begin{align}
\LL
& = \sum_{j=1}^N \left[
    \frac{\CT\dx}{2}\dot{\phi}_j{}^2 - \frac{(\phi_j-\phi_{j-1})^2}{2\LT\dx}
  \right]
+ \frac{\CC}{2}(\dot{\phi}_0-\dot{\varPhi}_0)^2
\nonumber \\ & \quad
+ \sum_{j=1}^{\infty} \left[
    \frac{\CT\dx}{2}\dot{\varPhi}_j{}^2 - \frac{(\varPhi_j-\varPhi_{j-1})^2}{2\LT\dx}
  \right]
+ \LA.
\end{align}
The canonical momenta are obtained as follows:
\begin{subequations}
\begin{align}
q_0
& = \frac{\partial L}{\partial\dot{\phi}_0}
  = \CC( \dot{\phi}_0 - \dot{\varPhi}_0 ), \\
q_j
& = \frac{\partial L}{\partial\dot{\phi}_j}
  = \CT\dx\dot{\phi}_j + \frac{\partial \LA}{\partial\dot{\phi}_j}, \\
Q_0
& = \frac{\partial L}{\partial\dot{\varPhi}_0}
  = \CC( \dot{\varPhi}_0 - \dot{\phi}_0 ), \\
Q_j
& = \frac{\partial L}{\partial\dot{\varPhi}_j}
  = \CT\dx\dot{\varPhi}_j.
\end{align}
\end{subequations}
In the continuous base, the Lagrange equations are derived as
\begin{subequations}
\begin{align}
\CC\left[ \ddot{\phi}(0^+) - \ddot{\varPhi}(0^-) \right]
& = \frac{1}{\LT}\left.\frac{\partial\phi(x)}{\partial x}\right|_{x=0^+}, \\
\CT\ddot{\phi}(x) + \frac{1}{\dx}\ddt{}\frac{\partial \LA}{\partial \dot{\phi}(x)}
& = \frac{1}{\LT}\ddxx{\phi(x)} + \frac{1}{\dx}\frac{\partial \LA}{\partial \phi(x)}, \\
\CC\left[ \ddot{\varPhi}(0^-) - \ddot{\phi}(0^+) \right]
& = - \frac{1}{\LT}\left.\frac{\partial\varPhi(x)}{\partial x}\right|_{x=0^-}, \\
\CT\ddot{\varPhi}(x)
& = \frac{1}{\LT}\ddxx{\varPhi(x)}.
\end{align}
\end{subequations}
Here, the motion of fields inside and outside the resonator are described
by the second and the last equation, respectively,
and the first and third equations are the boundary conditions.
From these equations, we get the following relation
concerning the incoming, outgoing, and internal fields at the capacitance $\CC$
($x = 0^+$):
\begin{align}&
\omega^2\CC\left[ \ophi^+(0^+,\omega) - \oPhiin^+(\omega) - \oPhiout^+(\omega) \right]
\nonumber \\ & \quad
= - \frac{\ii\omega}{\LT v}\left[ \oPhiin^+(\omega) - \oPhiout^+(\omega) \right].
\end{align}
Then, the outgoing field is expressed as
\begin{equation} \label{eq:oPhiout_TLR-TL} % !!!!!!!!!!!!!!!!!!!!!!!!!!!!!!!!!
\oPhiout^+(\omega) = \frac{\ophi^+(0^+,\omega) - [1-\ii\varLambda(\omega)]\oPhiin^+(\omega)}{1 + \ii\varLambda(\omega)}.
\end{equation}

On the other hand, from the Lagrangian or its Hamiltonian,
the equation of motion of $\osigma_{\mu,\nu}$ is approximately obtained as
\begin{align}
\ddt{}\osigma_{\mu,\nu}
& = - \ii\omega_{\nu,\mu}\osigma_{\mu,\nu}
+ \left[ \osigma_{\mu,\nu}, \ophi(0^+)\right]
  \int_0^{\infty}\dd\omega\ \ee^{-\ii\omega t}
    \frac{\ii\omega^2\CC}{\hbar}
\nonumber \\ & \quad \times
    \left[ \ophi^+(0^+,\omega) - \oPhiin^+(\omega) - \oPhiout^+(\omega) \right].
\end{align}
Substituting Eq.~\eqref{eq:oPhiout_TLR-TL} into this equation,
we get in the following equation in good-cavity ($\CC \gg \CT\cavlen$)
and independent-transition limit:
\begin{subequations}
\begin{align}
\ddt{}\osigma_{\mu,\nu}
& = - \ii\omega_{\nu,\mu}\osigma_{\mu,\nu}
+ \left[ \osigma_{\mu,\nu}, \ophi(0^+)\right]
  \int_0^{\infty}\dd\omega\ \ee^{-\ii\omega t}
\nonumber \\ & \quad \times
  \left[
    \frac{\omega^3\ZT\CC{}^2}{\hbar} \ophi^+(0^+,\omega)
  + \ii\sqrt{\frac{2\omega^3\ZT\CC{}^2}{\hbar}} \oain(\omega)
  \right].
\end{align}
On the other hand, Eq.~\eqref{eq:oPhiout_TLR-TL} is rewritten in the good cavity limit as
\begin{equation}
\oaout(\omega) \simeq \oain(\omega)
- \ii\sqrt{\frac{2\omega^3\ZT\CC{}^2}{\hbar}}
  \ophi^+(\omega).
\end{equation}
\end{subequations}
From these relations, we can conclude that the SEC is expressed as
\begin{equation}
\oHSE \simeq \int_0^{\infty}\dd\omega\ \hbar\sqrt{\frac{2\omega^3\ZT\CC{}^2}{2\pi\hbar}}
\left[ \oalphad(\omega) \ophi^+(0^+) + \Hc \right].
\end{equation}
When we define the loss rate of the $m$-th bare resonator mode
with frequency $\omega_m$ as
\begin{equation}
\kappaTLR_m = \frac{2\omega_m^2\ZT\CC{}^2}{\CT\cavlen},
\end{equation}
the SEC Hamiltonian is rewritten as
\begin{equation}
\oHSE
\simeq \int_0^{\infty}\dd\omega\
  \hbar\sqrt{\frac{\kappa_m}{2\pi}\left(\frac{\omega}{\omega_m}\right)^3\frac{\omega_m\cavlen}{\hbar v\ZT}}
  \left[ \oalphad(\omega) \ophi^+(0^+) + \Hc \right].
\end{equation}

% \bibliographystyle{bamba_notitle_href}
% \bibliography{../../../../SkyDrive/bib/list,../../../../SkyDrive/bib/bamba}

\end{document}